\documentclass[twocolumn,amsmath,nofootinbib]{revtex4} 
\usepackage{url}
\usepackage{amssymb}
\usepackage{amsfonts}
\usepackage{amsbsy}
\usepackage{graphicx}
\usepackage{hyperref}
\usepackage{array} 
\usepackage{multirow}
\usepackage{epsfig}
\usepackage{graphics}
\usepackage{amssymb}
\usepackage{amsmath}
\usepackage{comment}
\usepackage{enumerate}
\usepackage{subfigure}
\newcommand{\eqn}[1]{\begin{equation}\begin{split} #1 \end{split}\end{equation}}
\newcommand{\lp}{\left (}
\newcommand{\rp}{\right )}

\newcommand{\R}{\mathbb{R}}

\newcommand{\C}{\mathbb{C}}

\newcommand{\p}[2]{\frac{\partial #1}{\partial #2}}

\newcommand{\inv}{^{-1}}

\def\ie{{\frenchspacing\it i.e.}}
\def\eg{{\frenchspacing\it e.g.}}
\def\etc{{\frenchspacing\it etc.}}

\def\expec#1{\langle#1\rangle}

\def\A{\textbf{A}}
\def\B{\textbf{B}}
\def\C{\textbf{C}}

\def\H{\textbf{H}}

\def\M{\textbf{M}}
\def\N{\textbf{N}}
\def\W{\textbf{W}}
\def\vphi{\boldsymbol{\phi}}
\def\m{\boldsymbol{\mu}}
\def\ss{\boldsymbol{\sigma}}
\def\b{\textbf{b}}
\def\c{\textbf{c}}
\def\d{\textbf{d}}
\def\f{\textbf{f}}
\def\p{\textbf{p}}
\def\rvec{\textbf{r}}
\def\v{\textbf{v}}

\def\ex{y}
\def\wy{x}

\def\x{\textbf{\ex}}
\def\xh{\widehat{\ex}}
\def\y{\textbf{\wy}}

\def\omit#1{}

\def\spose#1{\hbox to 0pt{#1\hss}}
\def\simlt{\mathrel{\spose{\lower 3pt\hbox{$\mathchar"218$}}
   \raise 2.0pt\hbox{$\mathchar"13C$}}}
\def\simgt{\mathrel{\spose{\lower 3pt\hbox{$\mathchar"218$}}
     \raise 2.0pt\hbox{$\mathchar"13E$}}}
 \def\simpropto{\mathrel{\spose{\lower 3pt\hbox{$\mathchar"218$}}
     \raise 2.0pt\hbox{$\propto$}}}

\def\beq#1{\begin{equation}\label{#1}}
\def\eeq{\end{equation}}
\def\beqa#1{\begin{eqnarray}\label{#1}}
\def\eeqa{\end{eqnarray}}
\def\eq#1{equation~(\ref{#1})}	
\def\Eq#1{Equation~(\ref{#1})}

\def\fig#1{Figure~\ref{#1}}
\def\Fig#1{Figure~\ref{#1}}

\def\Sec#1{Section~\ref{#1}}

\parskip=\medskipamount
\begin{document}
\title{Why does deep and cheap learning work so well?\footnote{Published in {\it Journal of Statistical Physics}:\\ \url{https://link.springer.com/article/10.1007/s10955-017-1836-5}}}
%\title{Why is cheap learning possible?}
\author{Henry W. Lin, Max Tegmark, and David Rolnick}
\address{Dept.~of Physics, Harvard University, Cambridge, MA 02138}
\address{Dept.~of Physics, Massachusetts Institute of Technology, Cambridge, MA 02139}
\address{Dept.~of Mathematics, Massachusetts Institute of Technology, Cambridge, MA 02139}
\begin{abstract}
We show how the success of deep learning could depend not only on mathematics but also on physics:
although well-known mathematical theorems guarantee that neural networks can approximate arbitrary functions well, the class of functions of practical interest can frequently be approximated through ``cheap learning'' with exponentially fewer parameters than generic ones.
We explore how properties frequently encountered in physics such as symmetry, locality, compositionality, 
and polynomial log-probability translate into exceptionally simple neural networks. 
We further argue that when the statistical process generating the data is of a certain hierarchical form prevalent in physics and machine-learning, a deep neural network can be more efficient than a shallow one. We formalize these claims using information theory and discuss the relation to the renormalization group.
We prove various ``no-flattening theorems" showing when efficient linear deep networks cannot be accurately approximated by shallow ones without efficiency loss; for example, we show that $n$ variables cannot be multiplied using fewer than $2^n$ neurons in a single hidden layer.
\end{abstract}
\date{July 21 2017}
\vspace{10mm}	

\maketitle
%===============================================================
%===============================================================
%===============================================================

%=======================================================
\section{Introduction}
\label{IntroSec}
%=======================================================
Deep learning works remarkably well, and has helped dramatically improve the state-of-the-art in areas ranging from speech recognition, translation and visual object recognition to drug discovery, genomics and automatic game playing \cite{lecun2015deep,bengio2009}.
%handwriting recognition \cite{mnist}
%image captioning 
%game playing \cite{deepMindAtari}
However, it is still not fully understood {\it why} deep learning works so well. 
In contrast to GOFAI (``good old-fashioned AI") algorithms that are hand-crafted and fully understood analytically, many algorithms using artificial neural networks are understood only at a heuristic level, where we empirically know that certain training protocols employing large data sets will result in excellent performance. 
This is reminiscent of the situation with human brains: we know that if we train a child according to a certain curriculum, she will learn certain skills --- but we lack a deep understanding of how her brain accomplishes this.

This makes it timely and interesting to develop new analytic insights on deep learning and its successes, which is the goal of the present paper. Such improved understanding is not only interesting in its own right, and for potentially providing new clues about how brains work, but it may also have practical applications. Better understanding the shortcomings of deep learning may suggest ways of improving it, both to make it more capable and to make it more robust
\cite{russell2015research}.

\subsection{The swindle: why does ``cheap learning'' work?}

Throughout this paper, we will adopt a physics perspective on the problem, to prevent application-specific details from obscuring simple general results related to dynamics, symmetries, renormalization, {\etc}, and to exploit useful similarities between deep learning and statistical mechanics.
%One of the central problems in machine learning is function approximation:

\begin{figure}[phbt]
%\vskip-4cm
\centerline{\includegraphics[width=58mm]{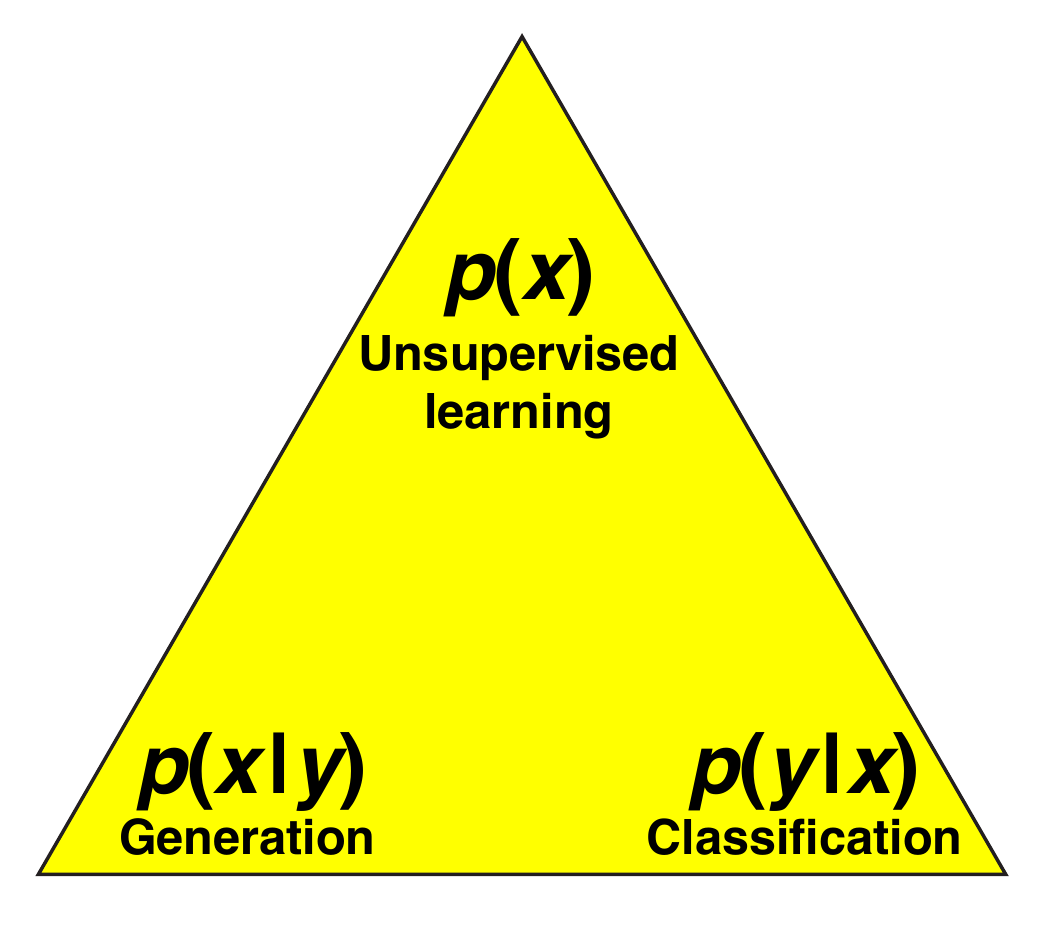}}
%\vskip-5.3cm
\caption{In this paper, we follow the machine learning convention where $\x$ refers to the model parameters and $\y$ refers to the data, thus viewing $\y$ as a stochastic function of $\x$ (please beware that this is the opposite of the common mathematical convention that $\x$ is a function of $\y$).
Computer scientists usually call $\y$ a {\it category} when it is discrete and a {\it parameter vector} when it is continuous.
Neural networks can approximate probability distributions. 
Given many samples of random vectors $\x$ and $\y$, unsupervised learning attempts to approximate the joint probability distribution of $\x$ and $\y$ without making any assumptions about causality. 
{\it Classification} involves estimating the probability distribution for $\x$ given $\y$.
The opposite operation, estimating the probability distribution of $\y$ given $\x$ is often called {\it prediction} when 
$\x$ causes $\y$ by being earlier data in a time sequence; in other cases where $\x$ causes $\y$, for example via a generative model, this operation is sometimes known as probability density estimation. Note that in machine learning, prediction is sometimes defined not as as outputting the probability distribution, but as sampling from it.
\label{TriangleFig}
}
\end{figure}

The task of approximating functions of many variables is central to most applications of machine learning, including unsupervised learning, classification and prediction, as illustrated in \fig{TriangleFig}.
% this covers most core 
For example, if we are interested in classifying faces, 
then we may want our neural network to implement a function where we feed in an image represented by a million greyscale pixels and get as output the probability distribution over a set of people that the image might represent. 

When investigating the quality of a neural net, there are several important factors to consider:
\begin{itemize}
\itemsep0mm
\item {\bf Expressibility:} What class of functions can the neural network express?
\item {\bf Efficiency:} How many resources (neurons, parameters, \etc) does the neural network require to approximate a given function? 
\item {\bf Learnability:} How rapidly can the neural network learn good parameters for approximating a function?
\end{itemize}
This paper is focused on expressibility and efficiency, and more specifically on the following well-known \cite{luck,shawe1998,poggio2015theory} problem:
{\it How can neural networks approximate functions well in practice, when the set of possible functions is exponentially larger than the set of practically possible networks?}
%We cannot even get past the first question before encountering a profound puzzle: how can neural networks possibly approximate functions well, when the set of possible functions is exponentially larger than the set of possible networks? 
For example, suppose that we wish to classify megapixel greyscale images into two categories, \eg, cats or dogs. 
If each pixel can take one of 256 values, then there are 
$256^{1000,000}$ 
possible images, and for each one, we wish to compute the probability that it depicts a cat.
This means that an arbitrary function is defined by a list of $256^{1000,000}$ probabilities, \ie, way more numbers than there are atoms in our universe (about $10^{78}$). 
%--- this is a bad analogy because we are comparing apples to oranges. If each atom is described by a small number of parameters, then the universe would be described by some doubly exponential large probability distribution/wave function.
% MT: The point we're making here is simply that this is too many parameters to be practically storable in our universe: even if we could write one parameter onto each atom, it wouldn't be enough. 

Yet neural networks with merely thousands or millions of parameters somehow manage to perform such classification tasks quite well. How can deep learning be so ``cheap'', in the sense of requiring so few parameters? 

We will see in below that neural networks perform a combinatorial swindle,  replacing exponentiation by multiplication: if there are say $n=10^6$ inputs taking $v=256$ values each, this swindle cuts the number of parameters from $v^n$ to $v\times n$ times some constant factor.
We will show that this success of this swindle depends fundamentally on physics: 
although neural networks only work well for an exponentially tiny fraction of all possible inputs, 
the laws of physics are such that the data sets we care about for machine learning (natural images, sounds, drawings, text, \etc) are also drawn from an exponentially tiny fraction of all imaginable data sets.
Moreover, we will see that these two tiny subsets are remarkably similar, enabling deep learning to work well in practice.

The rest of this paper is organized as follows. 
In \Sec{ShallowSec}, we present results for shallow neural networks with merely a handful of layers, focusing on simplifications due to locality, symmetry and polynomials.
In \Sec{DeepSec}, we study how increasing the depth of a neural network can provide polynomial or exponential efficiency gains even though it adds nothing in terms of expressivity, and we discuss the connections to renormalization, compositionality and complexity.
We summarize our conclusions in \Sec{ConclusionsSec}. %and discuss a technical point
%about renormalization and deep learning in \App{RenormalizationAppendix}.

%=======================================================
%\section{Expressibility: what functions can neural networks implement?}
\section{Expressibility and efficiency of shallow neural networks}
\label{ShallowSec}

Let us now explore what classes of probability distributions $p$ are the focus of physics and machine learning, and how accurately and efficiently neural networks can approximate them. We will be interested in probability distributions $p(\y|\ex)$, where $\y$ ranges over some sample space and $\ex$ will be interpreted either as another variable being conditioned on or as a model parameter.
%Although our results will be fully general, it will be useful to give concrete examples using the notation introduced in \fig{TriangleFig}. 
%of we give the mathematical notation from  concrete interpretations.
For a machine-learning example, we might interpret $\ex$ as an element of some set of animals $\{\text{cat},\text{dog},\text{rabbit},...\}$ and $\y$ as the vector of pixels in an image depicting such an animal, so that $p(\y|\ex)$ for $\ex=cat$ gives the probability distribution of images of cats with different coloring, size, posture, viewing angle, lighting condition, electronic camera noise, {\etc}
For a physics example, we might interpret $\ex$ as an element of some set of metals $\{\text{iron},\text{aluminum},\text{copper},...\}$ and $\y$ as the vector of magnetization values for different parts of a metal bar.
The prediction problem is then to evaluate $p(\y|\ex)$, whereas the classification problem is to evaluate $p(\ex|\y)$.

Because of  the above-mentioned ``swindle'', accurate approximations are only possible for a tiny subclass of all probability distributions. 
Fortunately, as we will explore below, the function $p(\y|\ex)$ often has many simplifying features enabling accurate approximation, because it follows from some simple physical law or some generative model with relatively few free parameters: for example, its dependence on $\y$ may exhibit symmetry, locality and/or be of a simple form such as the exponential of a low-order polynomial.  In contrast, the dependence of $p(\ex|\y)$ on $\ex$ tends to be more complicated; it makes no sense to speak of symmetries or polynomials involving a variable $\ex=cat$. 

Let us therefore start by tackling the more complicated case of modeling $p(\ex|\y)$.
%\subsection{Optimal classification with a neural networks}
This probability distribution $p(\ex|\y)$ is determined by the hopefully simpler function $p(\y|\ex)$ via Bayes' theorem:
\beq{BayesEq}
p(\ex|\y) = {p(\y|\ex)p(\ex)\over\sum_{\ex'} p(\y|\ex')(\ex')},
\eeq
where $p(\ex)$ is the probability distribution over $\ex$ (animals or metals, say) {\it a priori}, before examining the data vector $\y$.

%\subsection{Bayes theorem as a Boltzmann distribution}
\subsection{Probabilities and Hamiltonians}

It is useful to introduce the negative logarithms of two of these probabilities:
\beqa{HamiltonianDefEq}
H_\ex(\y)&\equiv& -\ln p(\y|\ex),\nonumber\\
\mu_\ex&\equiv& -\ln p(\ex).
\eeqa
Statisticians refer to $-\ln p$ as ``self-information'' or ``surprisal'', and statistical physicists refer to 
$H_\ex(\y)$ as the {\it Hamiltonian}, quantifying the energy of $\y$ (up to an arbitrary and irrelevant additive constant) given the parameter $\ex$.
Table~\ref{DictionaryTable} is a brief  dictionary translating between  physics and machine-learning terminology.
These definitions transform \eq{BayesEq} into the Boltzmann form 
\beq{BoltzmannEq}
p(\ex|\y) = {1\over N(\y)} e^{-[H_\ex(\y)+\mu_x]}, 
%p(x|\y) \propto e^{-[H_x(\y)+\mu_x]}
\eeq
where 
\beq{yDefEq}
N(\y)\equiv\sum_\ex e^{-[H_\ex(\y)+\mu_\ex]}.
\eeq
This recasting of \eq{BayesEq} is useful because the Hamiltonian tends to have properties making it simple to evaluate. We will see in \Sec{DeepSec} that it also helps understand the relation between deep learning and renormalization\cite{mehta}.

\begin{table}
{\footnotesize
\begin{tabular}{|l|l|}
\hline
Physics				&Machine learning\\
\hline
Hamiltonian			&Surprisal $-\ln p$\\
Simple $H$			&Cheap learning\\
Quadratic $H$			&Gaussian p\\
Locality				&Sparsity\\
Translationally symmetric $H$ &Convnet\\
Computing $p$ from $H$	&Softmaxing\\
Spin					&Bit\\
Free energy difference		&KL-divergence\\
Effective theory			&Nearly lossless data distillation\\ %�sufficient statistics"
Irrelevant operator		&Noise\\
Relevant operator		&Feature\\
\hline
\end{tabular}
\caption{Physics-ML dictionary.
\label{DictionaryTable}
}
}
\end{table}

\subsection{Bayes theorem as a softmax}

Since the variable $\ex$ takes one of a discrete set of values, we will often write it as an index instead of as an argument, as $p_\ex(\y)\equiv p(\ex|\y)$.
Moreover, we will often find it convenient to view all values indexed by $\ex$ as elements of a vector,  written in boldface, thus viewing $p_\ex$, $H_\ex$ and $\mu_\ex$
as elements of the vectors $\p$, $\H$ and $\m$, respectively. 
\Eq{BoltzmannEq} thus simplifies to
\beq{BoltzmannEq2}
\p(\y) = {1\over N(\y)} e^{-[\H(\y)+\m]}, 
\eeq
using the standard convention that a function (in this case $\exp$) applied to a vector acts on its elements.

We wish to investigate how well this vector-valued function $\p(\y)$ can be approximated by a neural net. 
A standard $n$-layer feedforward neural network maps vectors to vectors by applying a series of linear and nonlinear transformations in succession. 
Specifically, it implements vector-valued functions of the form 
\cite{lecun2015deep}
\beq{NNeq}
\f(\y)=\ss_n\A_n\cdots\ss_2\A_2\ss_1\A_1\y,
\eeq
where the  $\ss_i$ are relatively simple nonlinear operators on vectors and the $\A_i$ are affine transformations of the form 
$\A_i\y=\W_i\y+\b_i$ for matrices $\W_i$ and so-called bias vectors $\b_i$.
Popular choices for these nonlinear operators $\ss_i$ include
\begin{itemize}
\item {\it Local function}: apply some nonlinear function $\sigma$ to each vector element,
\item {\it Max-pooling}: compute the maximum of all vector elements,
\item {\it Softmax}: exponentiate all vector elements and normalize them to so sum to unity
\beq{SoftmaxDefEq}
\tilde{\ss}(\y)\equiv{e^{\y}\over\sum_ i e^{y_i}}.
\eeq
\end{itemize}
(We use $\tilde{\ss}$ to indicate the softmax function and $\ss$ to indicate an arbitrary non-linearity, optionally with certain regularity requirements).
%The softmax operator is therefore defined by 

This allows us to rewrite \eq{BoltzmannEq2} as
\beq{SoftmaxEq}
\p(\y)=\tilde{\ss}[-\H(\y)-\m].
\eeq
This means that if we can compute the Hamiltonian vector $\H(\y)$ with some $n$-layer neural net, we can evaluate the
desired classification probability vector $\p(\y)$ by simply adding a softmax layer. The $\m$-vector simply becomes the bias term in this final layer.

\subsection{What Hamiltonians can be approximated by feasible neural networks?}

It has long been known that  neural networks are universal\footnote{Neurons are universal analog computing modules in much the same way that NAND gates are universal digital computing modules: 
any computable function can be accurately evaluated by a sufficiently large network of them.
Just as NAND gates are not unique (NOR gates are also universal), nor is any particular neuron implementation --- indeed, {\it any} generic smooth nonlinear activation function is universal \cite{hornik1989multilayer,cybenko1989approximation}.}
approximators \cite{hornik1989multilayer,cybenko1989approximation}, in the sense that networks with virtually all popular nonlinear activation functions $\sigma(\wy)$ can approximate any smooth function to any desired accuracy --- even using merely a single hidden layer. However, these theorems do not guarantee that this can be accomplished with a network of feasible size, and the following simple example explains why they cannot:
There are $2^{2^n}$ different Boolean functions of $n$ variables, so a network implementing a generic function in this class 
requires at least $2^n$ bits to describe, \ie, more bits than there are atoms in our universe if $n>260$.

The fact that neural networks of feasible size are nonetheless so useful therefore implies that the class of functions we care about approximating is dramatically smaller.
We will see below in \Sec{PhysicsProbDistSec} that both physics and machine learning tend to favor Hamiltonians that are polynomials\footnote{
The class of functions that can be exactly expressed by a neural network must be invariant under composition, since adding more layers corresponds to using the output of one function as the input to another. Important such classes include linear functions, affine functions, piecewise linear functions (generated by the popular Rectified Linear unit ``ReLU'' activation function $\sigma(\wy) = \max[0,\wy]$), polynomials, continuous functions and smooth functions whose $n^{\rm th}$ derivatives are continuous. According to the Stone-Weierstrass theorem, both polynomials and piecewise linear functions can approximate continuous functions arbitrarily well.
} --- indeed, often ones that are sparse, symmetric and low-order. Let us therefore focus our initial investigation on Hamiltonians that can be expanded as a power series:
\beq{SeriesExpansionEq}
H_\ex(\y) = h+\sum_i h_i\wy_i + \sum_{i\le j} h_{ij}\wy_i\wy_j + \sum_{i\le j\le k} h_{ijk}\wy_i\wy_j\wy_k+ \cdots.
\eeq
If the vector $\y$ has $n$ components ($i=1,...,n$), then there are $(n+d)!/(n!d!)$ terms of degree up to $d$.

\subsubsection{Continuous input variables}

If we can accurately approximate multiplication using a small number of neurons, then we can construct a network efficiently approximating any polynomial 
$H_\ex(\y)$ by repeated multiplication and addition. We will now see that we can, using any smooth but otherwise arbitrary non-linearity $\sigma$ that is applied element-wise.  The popular logistic sigmoid activation function $\sigma(\wy)={1/(1+e^{-\wy})}$ will do the trick.

{\bf Theorem}: Let $\textbf{f}$ be a neural network of the form $\f=\A_2\ss\A_1$, 
where $\ss$ acts elementwise by applying some smooth non-linear function $\sigma$ to each element. Let the input layer, hidden layer and output layer have sizes
2, 4 and 1, respectively. Then $\f$ can approximate a multiplication gate arbitrarily well.

\begin{figure}[pt]
\centerline{\includegraphics[width=86mm]{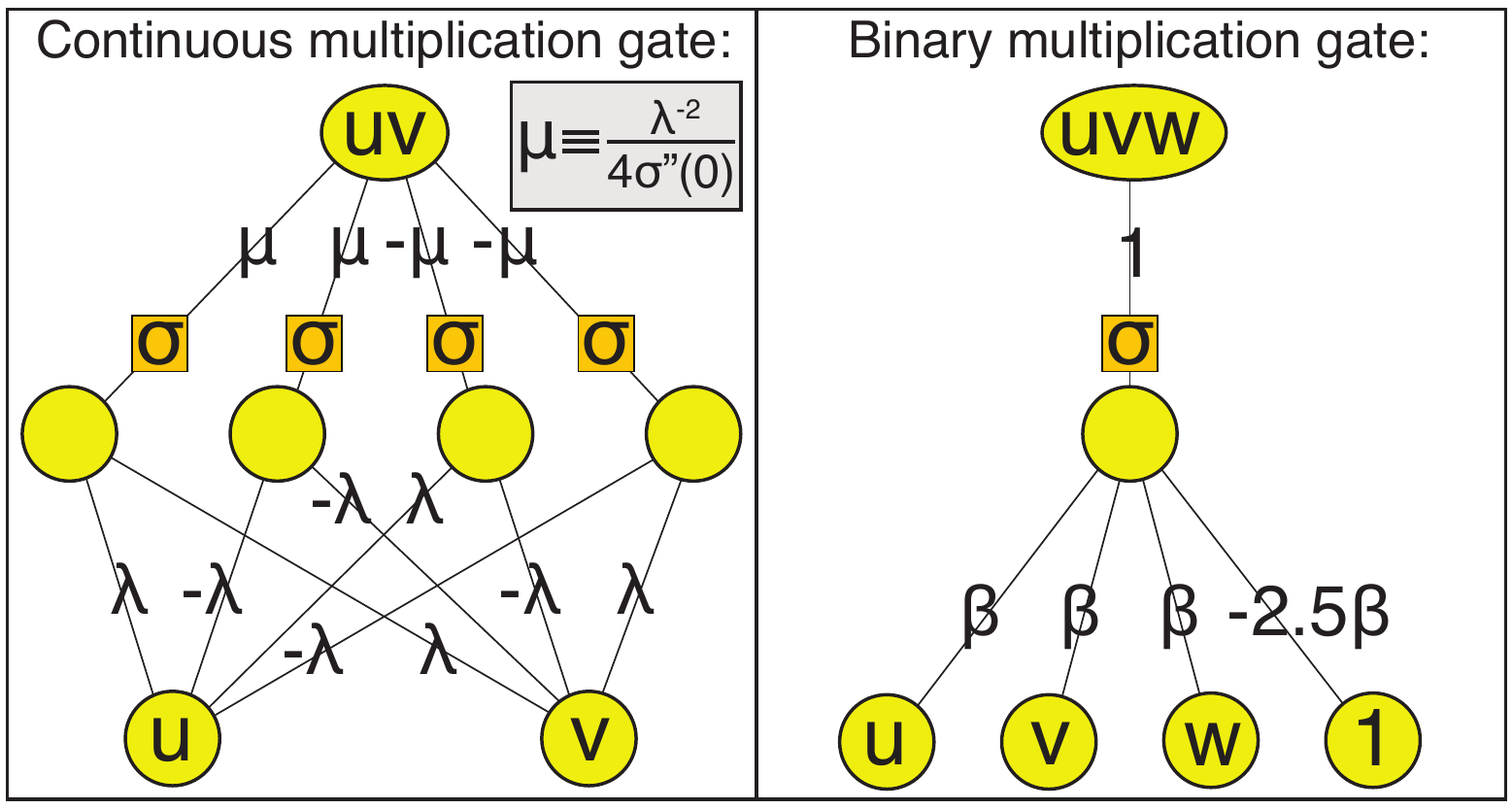}}
\caption{Multiplication can be efficiently implemented by simple neural nets, becoming arbitrarily accurate as $\lambda\to 0$ (left) and $\beta\to\infty$ (right). Squares apply the function $\sigma$, circles perform summation, and lines multiply by the constants labeling them. The ``1" input implements the bias term.
The left gate requires $\sigma''(0)\ne 0$, which can always be arranged by biasing the input to $\sigma$. The right gate requires the sigmoidal behavior $\sigma(\wy)\to 0$ and $\sigma(\wy)\to 1$ as $\wy\to-\infty$ and $\wy\to\infty$, respectively.
%$\sigma$
\label{MultiplicationFig}
}
\end{figure}

To see this, let us first Taylor-expand the function $\sigma$ around the origin:
\beq{gTaylorEq}
 \sigma(u) = \sigma_0 + \sigma_1 u + \sigma_2 {u^2\over 2} + \mathcal{O}(u^3).
 \eeq
Without loss of generality, we can assume that $ \sigma_2 \ne 0$: 
since $ \sigma$ is non-linear, it must have a non-zero second derivative at some point, so we can use the biases in $\A_1$ 
to shift the origin to this point to ensure $ \sigma_2\ne 0$.
%Without loss of generality, we can assume that $\sigma_0 = 0$ and $ \sigma_2 \ne 0$: if $ \sigma_0 \ne 0$, we can use the bias in $\textbf{A}_2$ to cancel $ \sigma_0$, and since $ \sigma$ is non-linear, it must have non-zero second derivative at some point, so we can use the biases in $\A_1$ to shift the origin to this point to ensure $ \sigma_2\ne 0$.
% THERE'S NO NEED TO ASSUME \sigma_0=0, SINCE IT CANCELS OUT :-)
% THE CUBIC TERMS CANCEL AS WELL!
\Eq{gTaylorEq} now implies that 
%\beq{xyApproxEq}
%{\scriptstyle g(x+y) + g(-x-y) - g(x-y) - g(-x+y)\over\scriptstyle 8 g_2} = x y + \mathcal{O}\left({\scriptstyle cubic\atop \scriptstyle terms}\right).
%\eeq
\beqa{xyApproxEq}
m(u,v)&\equiv&{\scriptstyle \sigma(u+v) +  \sigma(-u-v) -  \sigma(u-v) -  \sigma(-u+v)\over\scriptstyle 4\sigma_2}\nonumber\\
&=&uv \left[1 +  \mathcal{O}\left(u^2+v^2\right)\right], 
%\mathcal{O}\left({\scriptstyle cubic\atop \scriptstyle terms}\right), 
\eeqa
%Now we can check that for small $\ex$ and $y$, 
%\eqn{g(x+y) + g(-x-y) - g(x-y) - g(-x+y) \\ = 8 g_2 x y + \mathcal{O}(u^3).}
where we will term $m(u,v)$ the {\it multiplication approximator}.
Taylor's theorem guarantees that $m(u,v)$ is an arbitrarily good approximation of $uv$  for arbitrarily small $|u|$ and $|v|$. 
However, we can always make $|u|$ and $|v|$ arbitrarily small by scaling $\textbf{A}_1 \to \lambda \textbf{A}_1$ and then compensating by scaling $\textbf{A}_2 \to \lambda^{-2} \textbf{A}_2$. In the limit that $\lambda\to \infty$, this approximation becomes exact.
In other words, arbitrarily accurate multiplication can always be achieved using merely 4 neurons.
\Fig{MultiplicationFig} illustrates such a multiplication approximator. (Of course, a practical algorithm like stochastic gradient descent cannot achieve arbitrarily large weights, though a reasonably good approximation can be achieved already for $\lambda^{-1} \sim 10$.)

{\bf Corollary}: For any given multivariate polynomial and any tolerance $\epsilon >0$, there exists a neural network of fixed finite size $N$ (independent of $\epsilon$) that approximates the polynomial to accuracy better than $\epsilon$. Furthermore, $N$ is bounded by the complexity of the polynomial, scaling as
the number of multiplications required times a factor that is typically slightly larger than 4.\footnote{In addition to the four neurons required for each multiplication, 
additional neurons may be deployed to copy variables to higher layers bypassing the nonlinearity in $\sigma$. 
Such linear ``copy gates" implementing the function $u\to u$ are of course trivial to implement using a simpler version of
the above procedure: using $\A_1$ to shift and scale down the input to fall in a tiny range where $\sigma'(u)\ne 0$, and then scaling it up and shifting accordingly with $\A_2$.
}

This is a stronger statement than the classic universal universal approximation theorems for neural networks
\cite{hornik1989multilayer,cybenko1989approximation}, which guarantee that for every $\epsilon$ there exists some $N(\epsilon)$,
but allows for the possibility that  $N(\epsilon)\to\infty$ as $\epsilon \to 0$. An approximation theorem in \cite{pinkus1999approximation} provides an
$\epsilon$-independent bound on the size  of the neural network, but at the price of choosing a pathological function $\sigma$.

\subsubsection{Discrete input variables}

For the simple but important case where $\y$ is a vector of bits, so that $\wy_i=0$ or $\wy_i=1$, the fact that $\wy_i^2=\wy_i$ makes things even simpler.
This means that only terms where all variables are different need be included, which simplifies \eq{SeriesExpansionEq} to 
\beq{SeriesExpansionEq2}
H_\ex(\y) = h+\sum_i h_i \wy_i + \sum_{i< j} h_{ij} \wy_i\ex_j + \sum_{i< j<k} h_{ijk}\wy_i \wy_j \wy_k+ \cdots.
\eeq
The infinite series \eq{SeriesExpansionEq} thus gets replaced by a finite series with $2^n$ terms, ending with the
term  $h_{1...n}\wy_1\cdots\wy_n$. Since there are $2^n$ possible bit strings $\y$, the $2^n$ $h-$parameters in \eq{SeriesExpansionEq2} suffice to exactly parametrize an arbitrary function $H_\ex(\y)$.

The efficient multiplication approximator above multiplied only two variables at a time, thus requiring multiple layers to evaluate general polynomials. 
In contrast, $H(\y)$ for a bit vector $\y$ can be implemented using merely three layers as illustrated in \fig{MultiplicationFig}, where the middle layer evaluates the bit products and the third layer takes a linear combination of them. 
This is because bits allow an accurate multiplication approximator that takes the product of an arbitrary number of bits at once, exploiting the fact that a product of bits can be trivially determined from their sum: for example, the product $\wy_1\wy_2\wy_3=1$ if and only if the sum $\wy_1+\wy_2+\wy_3=3$.
This sum-checking can be implemented using one of the most popular choices for a nonlinear function $\sigma$:
the logistic sigmoid $\sigma(\wy)={1\over 1+e^{-\wy}}$
%\beq{SigmoidEq}
%\sigma(y)={1\over 1+e^{-y}},
%\eeq
which satisfies $\sigma(\wy)\approx 0$ for $\wy\ll 0$ and  $\sigma(\wy)\approx 1$ for $\wy\gg 1$.
To compute the product of some set of $k$ bits described by the set $K$ (for our example above, $K=\{1,2,3\}$), we let $\A_1$ and $\A_2$
shift and stretch the sigmoid to exploit the identity 
\beq{StretchedSigmoidEq}
\prod_{i \in K} \wy_i = \lim_{\beta \to \infty}\sigma\left[ -\beta \left(k-  \frac{1}{2}- \sum_{\wy\in K}  \, \wy_i\right)\right].
%\prod_{i \in K} y_i = \lim_{\beta \to \infty}\sigma\left[ \beta\sum_{y\in K}\, y_i  - \beta\left(k+\frac{1}{2}\right)\right].
\eeq
Since $\sigma$ decays exponentially fast toward $0$ or $1$ as $\beta$ is increased, modestly large $\beta$-values suffice in practice; if, for example, we want the correct answer to $D=10$ decimal places, we merely 
need $\beta>D\ln 10\approx  23$.
In summary, when $\y$ is a bit string, an {\it arbitrary} function $p_\ex(\y)$ can be evaluated by a simple 3-layer neural network: the middle layer uses sigmoid functions to compute the products from \eq{SeriesExpansionEq2}, and the top layer performs the sums from \eq{SeriesExpansionEq2} and the softmax from 
\eq{SoftmaxEq}.

%Since the Stone-Weierstrass theorem states that any continuous function on a compact domain can be approximated as closely as desired by a polynomial, this means that feedforward networks can approximate arbitrary continuous functions (in machine learning applications, the input vector always comes from a compact set since the input variables are bounded).
%{\bf WHAT SHOULD BE CITE HERE? THERE MUST BE LOTS OF SIMILAR RESULTS IN THE LITERATURE!}

% One of the most popular nonlinear functions for deep learning is the {\it rectified linear unit} (``ReLU''):
% \beq{ReLUeq}
% \sigma(y) = \max[0,y].
% \eeq
% It is easy to see that a neural network with ReLUs can generate all piecewise linear functions. The standard proof of the 
% Stone-Weierstrass theorem shows that these two can approximate continuous functions arbitrarily well.

% \beq{sigmoidTaylorEq}
% \sigma(y) = x-{x^3\over 3}+O(x^5)
% \eeq

\subsection{What Hamiltonians do we want to approximate?}
\label{PhysicsProbDistSec}

We have seen that polynomials can be accurately approximated by neural networks using a number of neurons scaling either as the number of multiplications required (for the continuous case) or as the number of terms (for the binary case).
But polynomials {\it per se} are no panacea: with binary input, all functions are polynomials, and with continuous input, there are $(n+d)!/(n!d!)$ coefficients in a generic polynomial of degree $d$ in $n$ variables, which easily becomes unmanageably large. 
We will now discuss situations in which exceptionally simple polynomials that are sparse, symmetric and/or low-order play a special role in physics and machine-learning.

\subsubsection{Low polynomial order}

The Hamiltonians that show up in physics are not random functions, but tend to be polynomials of very low order, typically of degree ranging from 2 to 4. The simplest example is of course the harmonic oscillator, which is described by a Hamiltonian that is quadratic in both position and momentum. 
There are many reasons why low order polynomials show up in physics. Two of the most important ones are that sometimes a phenomenon can be studied perturbatively, in which case, Taylor's theorem suggests that we can get away with a low order polynomial approximation. A second reason is renormalization: higher order terms in the Hamiltonian of a statistical field theory tend to be negligible if we only observe macroscopic variables.

%For reasons that are still not fully understood, our universe can be accurately described by polynomial Hamiltonians of low order $d$.
At a fundamental level, the Hamiltonian of the standard model of particle physics has $d=4$. There are many approximations of this quartic Hamiltonian that are accurate in specific regimes, for example 
the Maxwell equations governing electromagnetism, 
the Navier-Stokes equations governing fluid dynamics,
the Alv\'en equations governing magnetohydrodynamics and
various Ising models governing magnetization --- all of these approximations have Hamiltonians that are polynomials in the field variables, of degree $d$ ranging from 2 to 4.

This means that the number of polynomial coefficients in many examples is not infinite as in \eq{SeriesExpansionEq} or exponential in $n$ as in 
\eq{SeriesExpansionEq2}, merely of order $O(n^4)$.
%http://arxiv.org/pdf/1407.1035.pdf

There are additional reasons why we might expect low order polynomials. Thanks to the Central Limit Theorem \cite{clt}, many probability distributions in machine-learning and statistics can be accurately approximated by multivariate Gaussians, \ie,  of the form
\beq{GaussianEq}
p(\y)=e^{h + \sum_i h_j \wy_i - \sum_{ij} h_{ij} \wy_i \wy_j},
\eeq
which means that the Hamiltonian $H=-\ln p$ is a quadratic polynomial. More generally, the maximum-entropy probability distribution subject to constraints on some of the 
lowest moments, say expectation values of the form $\expec{\wy_1^{\alpha_1}\wy_2^{\alpha_2} \cdots\wy_n^{\alpha_n}}$ for some integers $\alpha_i \ge 0$ would lead to a Hamiltonian of degree no greater than $d\equiv \sum_i \alpha_i$ \cite{Jaynes}.
    
Image classification tasks often exploit invariance under translation, rotation, and various nonlinear deformations of the image plane that move pixels to new locations. All such spatial transformations are linear functions ($d=1$ polynomials) of the pixel vector $\y$. 
Functions implementing convolutions and Fourier transforms are also $d=1$ polynomials.

Of course, such arguments do not imply that we should expect to see low order polynomials in every application. If we consider some data set generated by a very simple Hamiltonian (say the Ising Hamiltonian), but then discard some of the random variables, the resulting distribution will in general become quite complicated. Similarly, if we do not observe the random variables directly, but observe some generic functions of the random variables, the result will generally be a mess. These arguments, however, might indicate that the probability of encountering a Hamiltonian described by a low-order polynomial in some application might be significantly higher than what one might expect from some naive prior. For example, a uniform prior on the space of all polynomials of degree $N$ would suggest that a randomly chosen polynomial would almost always have degree $N$, but this might be a bad prior for real-world applications.

We should also note that even if a Hamiltonian is described exactly by a low-order polynomial, we would not expect the corresponding neural network to reproduce a low-order polynomial Hamiltonian exactly in any practical scenario for a host of possible reasons including limited data, the requirement of infinite weights for infinite accuracy, and the failure of practical algorithms such as stochastic gradient descent to find the global minimum of a cost function in many scenarios. So looking at the weights of a neural network trained on actual data may not be a good indicator of whether or not the underlying Hamiltonian is a polynomial of low degree or not.
%So while the argument that a typical Hamiltonian encountered in ML problem is NOT a low-order polynomial is an important one, it cannot be conclusive.

\subsubsection{Locality}

One of the deepest principles of physics is {\it locality}: that things directly affect only what is in their immediate vicinity. 
When physical systems are simulated on a computer by discretizing space onto a rectangular lattice, locality manifests itself by allowing only nearest-neighbor interaction.
In other words, almost all coefficients in \eq{SeriesExpansionEq} are forced to vanish, and the total number of non-zero coefficients grows only linearly with $n$.
For the binary case of \eq{SeriesExpansionEq}, which applies to magnetizations (spins) that can take one of two values, locality also limits the degree $d$ to be no greater than the number of neighbors that a given spin is coupled to (since all variables in a polynomial term must be different).

Again, the applicability of these considerations to particular machine learning applications must be determined on a case by case basis. Certainly, an arbitrary transformation of a collection of local random variables will result in a non-local collection. (This might ruin locality in certain ensembles of images, for example). But there are certainly cases in physics where locality is still approximately preserved, for example in the simple block-spin renormalization group, spins are grouped into blocks, which are then treated as random variables. To a high degree of accuracy, these blocks are only coupled to their nearest neighbors.
Such locality is famously exploited by both biological and artificial visual systems, whose first neuronal layer performs merely fairly local operations.

%This can be stated more generally and precisely using the Markov network formalism \cite{markovnet}. View the spins as vertices of a Markov network; the edges represent dependencies. Let $N_c$ be the {\it clique cover number} of the network (the smallest number of cliques whose union is the entire network) and let $S_c$ be the size of the largest clique. Then the number of required neurons is $\le N_c 2^{S_c}$. For fixed $S_c$, $N_c$ is proportional to the number of vertices, so locality means that the number of neurons scales only linearly with the number of spins $n$.

\subsubsection{Symmetry}

Whenever the Hamiltonian obeys some symmetry (is invariant under some transformation), the number of independent parameters required to 
describe it is further reduced. For instance, many probability distributions in both physics and machine learning are invariant under translation and rotation. As an example, consider a vector $\y$ of air pressures $y_i$ measured by a microphone at times $i=1,...,n$.
Assuming that the Hamiltonian describing it has $d=2$ reduces the number of parameters $N$ from
$\infty$ to $(n+1)(n+2)/2$. Further assuming locality (nearest-neighbor couplings only) reduces this to $N=2n$, after which
requiring translational symmetry reduces the parameter count to $N=3$.
%Assuming that the Hamiltonian describing it has $d=2$ reduces the number of parameters $N$ from
%$N=\infty$ to $N=(n+1)(n+2)/2$: $H=h+\sum h_i y_i + \sum_{ij} h_{ij} y^i y^j$. 
%Further assuming locality (nearest-neighbor couplings only) reduces this to $N=2n$:
%$H=h +\sum_{i=1}^n  h_i y_i +\sum_{i=2}^n  g_{i-1} y_i$.
%Finally, requiring translational symmetry reduces the parameter count to $N=3$, since all the $h_i$ must be equal and all the $g_i$ must be equal.
Taken together, the constraints on locality, symmetry and polynomial order reduce the number of continuous parameters in the Hamiltonian of the standard model of physics to merely 32 \cite{axions}.

Symmetry can reduce not merely the parameter count, but also the computational complexity. 
For example, if a linear 
vector-valued function $\f(\y)$ mapping a set of $n$ variables onto itself happens to satisfy translational symmetry, then it is a convolution (implementable by a convolutional neural net; ``convnet"),  which means that it can be computed with $n \log_2 n$ rather than $n^2$ multiplications using Fast Fourier transform.

%=======================================================
\section{Why deep?}
\label{DeepSec}
%======================================================

%* classes of functions closed under compositionality (also cite Bengio paper 1 [ref in poggio] for number of linear regions for RELU network?)

%* cite Pinkus theorem (cite Poggio refs [4-5])

Above we investigated how probability distributions from physics and computer science applications lent themselves to ``cheap learning", 
being accurately and efficiently approximated by neural networks with merely a handful of layers.
Let us now turn to the separate question of depth, \ie, the success of deep learning: 
what properties of real-world probability distributions  
cause efficiency to further improve when networks are made deeper?
This question has been extensively studied from a mathematical point of view \cite{sumproduct,poggio2016,mhaskarPoggio2016}, but mathematics alone cannot fully answer it, because part of the answer involves physics.
We will argue that the answer involves the hierarchical/compositional structure
of generative processes together with inability to efficiently ``flatten'' neural networks reflecting this structure.

% * This connects with compositionality, scale-invariance and renormalization.

\subsection{Hierarchical processess}
%\subsection{Generative hierarchy}

One of the most striking features of the physical world is its hierarchical structure. 
Spatially, it is an object hierarchy: elementary particles form atoms which in turn form molecules, cells, organisms, planets, solar systems, galaxies, {\etc} 
Causally, complex structures are frequently created through a distinct sequence of simpler steps.

\begin{figure*}[phbt]
\centerline{\includegraphics[width=150mm]{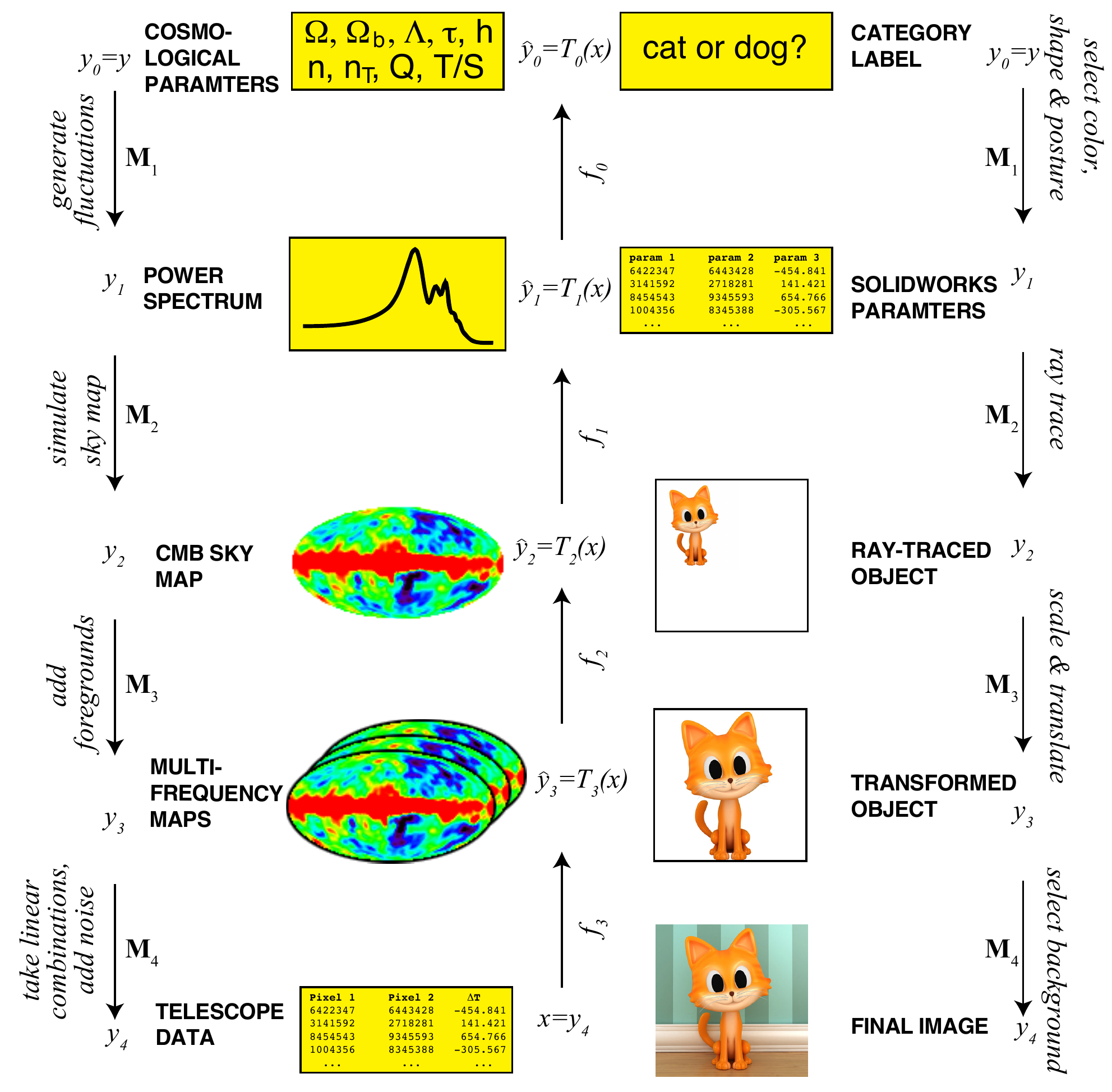}}
\caption{Causal hierarchy examples relevant to physics (left) and image classification (right). 
As information flows down the hierarchy $\ex_0\to \ex_1\to...\to \ex_n=y$, some of it is destroyed by random Markov processes.
However, no further information is lost as information flows optimally back up the hierarchy as $\xh_{n-1}\to...\to \xh_0$.
The right example is deliberately contrived and over-simplified for pedagogy; for example, translation and scaling are 
more naturally performed before ray tracing, which in turn breaks down into multiple steps.
\label{HierarchyFig}
}
\end{figure*}

\Fig{HierarchyFig} gives two examples of such causal hierarchies generating data vectors $\ex_0\mapsto \ex_1\mapsto ...\mapsto \ex_n$ 
that are relevant to physics and image classification, respectively. 
Both examples involve a Markov chain\footnote{If the next step in the generative hierarchy requires knowledge of not merely of the present state but also information of the past, the present state can be redefined to include also this information, thus ensuring that the generative process is a Markov process.}
where the probability distribution
$p(\ex_i)$ at the $i^{\rm th}$ level of the hierarchy is determined from its causal predecessor alone:
\beq{MarkovDefEq}
\p_i = \M_i\p_{i-1},
\eeq
where the probability vector $\p_i$ specifies the probability distribution of $p(\ex_i)$ according to $(\p_i)_\ex\equiv p(\ex_i)$ and the Markov matrix $\M_i$  specifies the transition probabilities between two neighboring levels, $p(\ex_{i}|\ex_{i-1})$.
Iterating \eq{MarkovDefEq} gives
\beq{MarkovEq2}
\p_n = \M_n \M_{n-1} \cdots \M_1 \p_0,
\eeq
so we can write the combined effect of the the entire generative process as a matrix product.

In our physics example (\fig{HierarchyFig}, left), a set of cosmological parameters $\ex_0$ (the density of dark matter, \etc) determines the power spectrum $\ex_1$ of density fluctuations in our universe, which in turn determines the pattern of cosmic microwave background radiation $\ex_2$ reaching us from our early universe, which gets combined with foreground radio noise from our Galaxy to produce the frequency-dependent sky maps ($\ex_3$) that are recorded by a satellite-based telescope that measures linear combinations of different sky signals and adds electronic receiver noise. For the recent example of the Planck Satellite \cite{Planck2015Results}, these datasets $\ex_i$, $\ex_2, ...$ contained about 
$10^1$, $10^4$, $10^8$, $10^9$ and $10^{12}$ numbers, respectively. 

More generally, if a given data set is generated by a (classical) statistical physics process, it must be described by an equation in the form of \eq{MarkovEq2}, since dynamics in classical physics is fundamentally Markovian: classical equations of motion are always first order differential equations in the Hamiltonian formalism. This technically covers essentially all data of interest in the machine learning community, although the fundamental Markovian nature of the generative process of the data may be an  in-efficient description.

Our toy image classification example (\fig{HierarchyFig}, right) is deliberately  contrived and over-simplified for pedagogy: $\ex_0$ is a single bit signifying ``cat or dog", which determines a set of parameters determining the animal's coloration, body shape, posture, {\etc} using approxiate probability distributions, which determine a 2D image via ray-tracing, which is scaled and translated by random amounts before a randomly generated background is added.

In both examples, the goal is to reverse this generative hierarchy to learn about
the input $\ex\equiv \ex_0$ from the output $\ex_n\equiv\wy$, specifically to provide the best possibile estimate of the probability distribution $p(\ex|y)=p(\ex_0|\ex_n)$ ---
\ie, to determine the probability distribution for the cosmological parameters and to determine the probability that the image is a cat, respectively.

\subsection{Resolving the swindle}

This decomposition of the generative process into a hierarchy of simpler steps
helps resolve the``swindle" paradox from the introduction:
although the number of parameters required to describe an arbitrary function of the 
input data $y$ is beyond astronomical, the generative process can be specified by a more modest number of parameters, because each of its steps can.
Whereas specifying an arbitrary probability distribution over multi-megapixel images $\wy$ requires far more bits than there are atoms in our universe, 
the information specifying how to compute the probability distribution $p(\wy|\ex)$ for a microwave background map fits into a handful of published journal articles or software packages \cite{cmbfast,cl,BJK,mapforegs,PlanckForegs,mapmaking,
WMAPmapmaking}.
For a megapixel image of a galaxy, its entire probability distribution is defined by the standard model of particle physics with its 32 parameters \cite{axions}, which together specify the process transforming primordial hydrogen gas into galaxies. 

The same parameter-counting argument can also be applied to all artificial images of interest to machine learning: for example, giving the simple low-information-content instruction {\it ``draw a cute kitten"} to a random sample of artists will produce a wide variety of images $y$ with a complicated probability
distribution over colors, postures, {\etc}, as each artist makes random choices at a series of steps. Even the pre-stored information about cat probabilities in these artists' brains is modest in size.

Note that a random resulting image typically contains much more information than 
the generative process creating it; for example, the simple instruction {\it ``generate a random string of $10^{9}$ bits"} contains much fewer than $10^{9}$ bits.
Not only are the typical steps in the generative hierarchy specified by a non-astronomical number of parameters, but
as discussed in  \Sec{PhysicsProbDistSec}, it is plausible that neural networks can implement each of the steps 
efficiently.\footnote{Although our discussion is focused on describing probability distributions, which are not random, 
stochastic neural networks can generate random variables as well.
In biology, spiking neurons provide a good random number generator, and in machine learning, stochastic architectures such as restricted Boltzmann machines \cite{hinton2010practical} do the same.}
%As we saw in  \Sec{ShallowSec}, it is plausible that each of the generative steps in the hierarchy can be implemented by a relatively simple neural net (when including random noise as needed;  in biology, spiking neurons provide a good random number generator, and in machine learning, stochastic architectures such as restricted Boltzmann machines \cite{} do the same).
 
A deep neural network stacking these simpler networks on top of one another would then implement the
entire generative process efficiently.
In summary, the data sets and functions we care about form a minuscule minority, and it is plausible 
that they can also be efficiently implemented by neural networks reflecting their generative process.
So what is the remainder? Which are the data sets and functions that we do {\it not} care about?

Almost all images are indistinguishable from random noise, and almost all data sets and functions are indistinguishable from completely random ones.
This follows from Borel's theorem on normal numbers \cite{Borel1909}, which states
that almost all real numbers have a string of decimals that would pass any randomness test, \ie, are indistinguishable from random noise.
Simple parameter counting shows that deep learning (and our human brains, for that matter) would fail to implement almost all such functions, and training would fail to find any useful patterns. To thwart pattern-finding efforts.
cryptography therefore aims to produces random-looking patterns.
Although we might expect the Hamiltonians describing human-generated data sets such as drawings, text and music to be more complex than those describing simple physical systems, we should nonetheless expect them to resemble the natural
data sets that inspired their creation much more than they resemble random functions.

\subsection{Sufficient statistics and hierarchies}

The goal of deep learning classifiers is to reverse the hierarchical generative process as well as possible, 
to make inferences about the input $\ex$ from the output $\wy$.
Let us now treat this hierarchical problem more rigorously using information theory.

%\eqn{\p = \M_n \M_{n-1} \cdots \M_1 \p_0 }

%* define and explain sufficient statistics and relation to the hierarchy figure,
   %and explain how insufficient (but nearly sufficient) statistics are still very useful
 Given $P(\ex|\wy)$, a {\it sufficient statistic} $T(\wy)$ is defined by the equation $P(\ex|\wy) = P(\ex|T(\wy))$ and has played an important role in statistics for almost a century \cite{fisher}. All the information about $\ex$ contained in $\wy$ is contained in the sufficient statistic. A {\it minimal sufficient statistic} \cite{fisher} is some sufficient statistic $T_*$ which is a sufficient statistic for all other sufficient statistics. This means that if $T(y)$ is sufficient, then there exists some function $f$ such that $T_*(y) = f (T(y))$. 
As illustrated in \fig{HierarchyFig},  $T_*$ can be thought of as a an information distiller, optimally compressing the data so as to retain all information relevant to determining $\ex$ and discarding all irrelevant information.
%In the language of neuroscience, $T(y)$ can be thought of as an invariant 

%Let us return to the case where our model is produced by a Markovian generative process  $X_0 \to X_1 \to \cdots \to X_n$ and see what this implies about the form of the sufficient statistic. We will think of $x \in X_0$ as the categories in our classification problem and $y \in X_n$ as the data that we feed in to the neural network. The Markov assumption means that if we define a vector $\p$ whose components are $\p_y = P(y)$, then we can write
%\beq{MarkovEq2}
%\p = \M_n \M_{n-1} \cdots \M_1 \p_0,
%\eeq
%where $(\p_0)_x = P(x)$, and each $\M_i$ is the Markov matrix giving the conditional probabilities $P(x_{i}|x_{i-1})$. We are now ready to state simple but important results.

%Now consider the minimal sufficient statistic of $P(x_i|y)$, which we call $T_i(y)$. Amazingly, $T_{i-1}(y)$ is also a sufficient statistic for $P(x_i|y)$ given that the generative model is Markovian.%, since 
%\eqn{P(x_i|y) &= \sum_{x_{i-1}}  P(x_i|x_{i-1})P(x_{i-1}|y)\\
%&= \sum_{x_{i-1}}  P(x_i|x_{i-1})P\lp x_{i-1}|T_{i-1}(y)\rp\\
%&= P\lp x_i|T_{i-1}(y)\rp,
%}
%Now since $T_i$ is minimal, for $i>1$ we must have $T_i(y) = f_i \circ T_{i-1}$ for some $f_i$. Defining $f_1 \equiv T_1(y)$, we obtain
%\eqn{T_*(y) = \lp f_1 \circ f_{2} \circ \cdots \circ f_n\rp (y).}

The sufficient statistic formalism enables us to state some simple but important results that apply to any hierarchical 
generative process cast in the Markov chain form of \eq{MarkovEq2}.

{\bf Theorem 2}: 
%Let $X_0 \to X_1 \to \cdots \to X_n$ be a Markov chain. 
%Let $T_i:X_n \to \hat{X}_i$ be a minimal sufficient statistic of $P(x_i|x_n)$. Then for $i\ge 0$, there exists some $f_{i}$ such that $T_i = f_{i} \circ T_{i+1}$.
Given a Markov chain described by our notation above, let $T_i$ be a minimal sufficient statistic of $P(\ex_i|\ex_n)$. Then there exists some functions $f_{i}$ such that $T_i = f_{i} \circ T_{i+1}$.
%x_n)$. Then there exists some functions $f_{i}$ such that $T_i = f_{i} \circ T_{i+1}$:
%
% zzz HENRY: ALAS, I CAN'T GET THIS LATEX PACKAGE TO INSTALL:
%\[
%\begin{tikzcd}[column sep=small]
%& x_n \arrow[dl, "T_{i+1}"'] \arrow[dr,"T_{i}"]     & \\
%\hat{x}_{i+1} \arrow[rr, dashed, "f_{i}"] &                          & \hat{x}_{i}
%\end{tikzcd}
%\]
%
More casually speaking, the generative hierarchy of \fig{HierarchyFig} can be optimally reversed one step at a time: there are functions $f_i$ that optimally undo each of the steps, distilling out all information about the level above that was not destroyed by the Markov process.
Here is the proof. Note that for any $k\ge 1$, the ``backwards''  Markov property $P(\ex_i |\ex_{i+1},\ex_{i+k}) = P(\ex_i|\ex_{i+1})$ follows from the Markov property via Bayes' theorem:
\eqn{P(\ex_i |\ex_{i+k}, \ex_{i+1}) &=
% \frac{P(x_{i+2} x_{i+1}|x_i) P(x_i)}{P(x_{i+2} x_{i+1}) } \\ 
 %&=
 \frac{P(\ex_{i+k}| \ex_i, \ex_{i+1}) P(\ex_i|\ex_{i+1})}{P(\ex_{i+k}| \ex_{i+1})} \\
  &= \frac{P(\ex_{i+k}| \ex_{i+1}) P(\ex_i|\ex_{i+1})}{P(\ex_{i+k}| \ex_{i+1})} \\
   &= P(\ex_{i}|\ex_{i+1}).
}
Using this fact, we see that
\eqn{P(\ex_i|\ex_n) &= \sum_{\ex_{i+1}} P(\ex_i|\ex_{i+1} \ex_n) P(\ex_{i+1}|\ex_n) \\
&= \sum_{\ex_{i+1}} P(\ex_i|\ex_{i+1}) P(\ex_{i+1}|T_{i+1} (\ex_n) ). \\
}
Since the above equation depends on $\ex_n$ only through $T_{i+1}(\ex_n)$, this means that $T_{i+1}$ is a sufficient statistic for $P(\ex_i|\ex_n)$. But since $T_i$ is the minimal sufficient statistic, there exists a function $f_{i}$ such that $T_i = f_{i} \circ T_{i+1}$.

{\bf Corollary 2}: With the same assumptions and notation as theorem 2, define the function $f_0(T_0) = P(\ex_0|T_0)$ and $f_n = T_{n-1}$. Then
\eqn{P(\ex_0|\ex_n) = \lp f_0 \circ f_1 \circ \cdots \circ f_n \rp (\ex_n).}
The proof is easy. By induction,
\eqn{\label{comp}T_0 = f_1 \circ f_2 \circ \cdots \circ T_{n-1},}
which implies the corollary.

Roughly speaking, Corollary 2 states that {\it the structure of the inference problem reflects the structure of the generative process}.
In this case, we see that the neural network trying to approximate $P(\ex|\wy)$ must approximate a compositional function. We will argue below in \Sec{NoFlatteningSec} that in many cases, this can only be accomplished efficiently if the neural network has $\gtrsim n$ hidden layers.

In neuroscience parlance, the functions $f_i$ compress the data into forms
with ever more {\it invariance} \cite{riesenhuber2000models}, containing features invariant under irrelevant 
transformations (for example background substitution, scaling and translation).

Let us denote the distilled vectors $\xh_i\equiv f_i(\xh_{i+1})$, where $\xh_n\equiv y$.
As summarized by \fig{HierarchyFig}, as information flows down the hierarchy $\ex=\ex_0\to \ex_1\to...\\ex_n=\wy$, some of it is destroyed by random processes.
However, no further information is lost as information flows optimally back up the hierarchy as $y\to\xh_{n-1}\to...\to \xh_0$.

\subsection{Approximate information distillation}

Although minimal sufficient statistics are often difficult to calculate in practice, it is frequently possible to come up with statistics which are nearly sufficient in a certain sense which we now explain.

An equivalent characterization of a sufficient statistic is provided by information theory \cite{kldiv,coverthomas}. The {\it data processing inequality} \cite{coverthomas} states that for any function $f$ and any random variables $x,y$, 
\beq{DataProcessingInequality}
I(x,y) \ge I(x,f(y)),
\eeq
where $I$ is the {\it mutual information}:
\eqn{I(x,y) = \sum_{x,y} p(x,y) \log \frac{p(x,y)}{p(x) p(y)}.}
A sufficient statistic $T(\wy)$ is a function $f(\wy)$ for which ``$\ge$" gets replaced by ``$=$" in \eq{DataProcessingInequality}, 
\ie, a function retaining all the information about $\ex$.

Even information distillation functions $f$ that are not strictly sufficient can be very useful as long as they distill out {\it most} of the relevant information and are computationally efficient. For example, it may be possible to trade some loss of mutual information with a dramatic reduction in the complexity of the Hamiltonian; e.g., $H_\ex\lp f(\wy)\rp$ may be considerably easier to implement in a neural network than $H_\ex \lp \wy \rp$.
Precisely this situation applies to the physical example described in \fig{HierarchyFig},
where a hierarchy of efficient near-perfect information distillers $f_i$ have been found,
the numerical cost of $f_3$ \cite{mapmaking,WMAPmapmaking}, $f_2$ \cite{mapforegs,PlanckForegs}, $f_1$ \cite{cl,BJK} and  $f_0$  \cite{Planck2015Results} scaling with the number of inputs parameters $n$
as $\mathcal{O}(n)$, $\mathcal{O}(n^{3/2})$, $\mathcal{O}(n^{2})$ and $\mathcal{O}(n^{3})$, respectively. More abstractly, the procedure of renormalization, ubiquitous in statistical physics, can be viewed as a special case of approximate information distillation, as we will now describe.

\subsection{Distillation and renormalization}

The systematic framework for distilling out desired information from unwanted ``noise'' in physical theories is known as Effective Field Theory \cite{kardar}. Typically, the desired information involves relatively large-scale features that can be experimentally measured, whereas the noise involves unobserved microscopic scales.  A key part of this framework is known as the {\it renormalization group} (RG) transformation \cite{kardar, cardy1996}. Although the connection between RG and machine learning has been studied or alluded to repeatedly \cite{johnson07, beny13, Saremi19022013, mehta, stoud16},  there are significant misconceptions in the literature concerning the connection which we will now attempt to clear up.

Let us first review a standard working definition of what renormalization is in the context of statistical physics, 
involving three ingredients: a vector $y$ of random variables, a course-graining operation $R$ and a requirement that this operation leaves the Hamiltonian invariant except for parameter changes. We think of $y$ as the microscopic degrees of freedom --- typically physical quantities defined at a lattice of points (pixels or voxels) in space. Its probability distribution is specified by a Hamiltonian $H_\ex(\wy)$, with some parameter vector $\ex$. We interpret the map $R: y\to y$ as implementing a coarse-graining\footnote{A typical renormalization scheme for a lattice system involves replacing many spins (bits) with a single spin according to some rule. In this case, it might seem that the map $R$ could not possibly map its domain onto itself, 
since there are fewer degrees of freedom after the coarse-graining. On the other hand, if we let the domain and range of $R$ differ, we cannot easily talk about the Hamiltonian as having the same functional form, since the renormalized Hamiltonian would have a different domain than the original Hamiltonian. Physicists get around this by taking the limit where the lattice is infinitely large, so that $R$ maps an infinite lattice to an infinite lattice.}
of the system. The random variable $R(y)$ also has a Hamiltonian, denoted $H'(R(y))$, which we require to have the same functional form as the original Hamiltonian $H_\ex$, although the parameters $\ex$ may change. In other words, $H'(R(\wy)) = H_{r(\ex)}(R(\wy))$ for some function $r$. Since the domain and the range of $R$ coincide, this map $R$ can be iterated $n$ times $R^n = R \circ R \circ \cdots R$, giving a Hamiltonian $H_{r^n(\ex)}(R^n(\wy))$ for the repeatedly renormalized data. Similar to the case of sufficient statistics, $P(\ex|R^n(\wy))$ will then be a compositional function.

Contrary to some claims in the literature, effective field theory and  the renormalization group have little to do with the idea of unsupervised learning and pattern-finding. Instead, the standard renormalization procedures in statistical physics are essentially a feature extractor for {\it supervised} learning, where the features typically correspond to long-wavelength/macroscopic degrees of freedom. In other words, effective field theory only makes sense if we specify what features we are interested in.
For example, if we are given data $\wy$ about the position and momenta of particles inside a mole of some liquid and is tasked with predicting from this data whether or not Alice will burn her finger when touching the liquid, a (nearly) sufficient statistic is simply the temperature of the object, which can in turn be obtained from some very coarse-grained degrees of freedom (for example, one could use the fluid approximation instead of working directly from the positions and momenta of $\sim 10^{23}$ particles). But without specifying that we wish to predict (long-wavelength physics), there is nothing natural about an effective field theory approximation.

%Note that there are several ways in which the RG differs from machine learning. 
%There are several reasons why RG is studied in statistical physics. In statistical physics, an important use of the RG is to efficiently numerically evaluating partition function $Z$, which is usually a function of macroscopic variables like temperature or pressure. Once the partition function is evaluated 

%Perhaps the reason that is closest to machine learning is the following: given that we only want to observe certain macroscopic quantities like energy, magnetization, etc., 
%First, the fixed points of the RG map $R(\alpha) = \alpha$ can signal a phase transition. Second, RG can be used to construct a so-called effective field theory from a microscopic theory. If $R$ corresponds to integrating out high energy (short-wavelength) information, $R^n(Y)$ can be thought of as {\it effective} degrees of freedom, e.g., the degrees of freedom that we actually care about in experiments.
%; we develop this connection further and clear up misconceptions found in previous literature in Appendix A. 

%Note that this map from $s' = f(s)$ can be iterated many times, so we could write 
%\eqn{s^{(n)} = f^n s.} 
%This means that although the renormalization group gives rise to insufficient statistics, the statistics still have the form given by equation (\ref{comp}).

%To obtain a more quantitative link between renormalization and deep-learning-style feature extraction, 
To be more explicit about the link between renormalization and deep-learning, consider a toy model for natural images. Each image is described by an intensity field $\phi(\rvec)$, where $\rvec$ is a 2-dimensional vector. We assume that an ensemble of images can be described by a quadratic Hamiltonian of the form
\beq{GaussianFieldEq}
H_\ex(\phi) = \int  \left[\ex_0 \phi^2 + \ex_1 (\nabla\phi)^2 + \ex_2 \lp \nabla^2\phi \rp^2 + \cdots \right]d^2 r.
\eeq
Each parameter vector $\ex$ defines an ensemble of images; we could imagine that the fictitious classes of images that we are trying to distinguish are all generated by Hamiltonians $H_\ex$ with the same above form but different parameter vectors $\ex$.
%We allow all terms which are quadratic in $y$ to be in the Hamiltonian, with each coefficient labeled by the number of derivatives involved, e.g., $x_7$ is followed by $(\nabla^7 y)^2$.
We further assume that the function $\phi(\rvec)$ is specified on pixels that are sufficiently close that derivatives can be well-approximated by differences. Derivatives are linear operations, so they can be implemented in the first layer of a neural network. The translational symmetry of \eq{GaussianFieldEq} allows it to be implemented with a convnet.
%This Hamiltonian involves computing arbitrarily many terms.
%The $R$-map is defined by $y' = R(y)$: 
%\beq{RdefEq}
%y'(\rvec/b) = b^{-4} \int d^2 y'\, W(\rvec-\rvec') y(\rvec'),
%\eeq
%where $W$ is some convolution kernel which has a width $b \times b$. This coarse-graining operation is essentially what you get by shrinking the image so that it can be viewed on larger scales.
%Under this transformation, it can be shown \cite{kardar} that
If can be shown \cite{kardar} that for any course-graining operation that replaces each block of $b\times b$ pixels by its average and divides the result by $b^2$, the Hamiltonian retains the form of \eq{GaussianFieldEq} but with the parameters $\ex_i$ replaced by
\beq{xRGeq}
\ex_i' = b^{2 - 2i} \ex_i.
\eeq
This means that all parameters $\ex_i$ with  $i\ge 2$ decay exponentially with $b$ as we repeatedly renormalize and $b$ keeps increasing,
so that for modest $b$, one can neglect all but the first few $\ex_i$'s. What would have taken an arbitrarily large neural network can now be computed on a neural network of finite and bounded size, assuming that we are only interested in classifying the data based only on the coarse-grained variables. These insufficient statistics will still have discriminatory power if we are only interested in discriminating Hamiltonians which all differ in their first few $C_k$.
In this example, the parameters $\ex_0$ and $\ex_1$ correspond to ``relevant operators" by physicists and ``signal" by machine-learners, 
whereas the remaining parameters correspond to ``irrelevant operators" by physicists and ``noise" by machine-learners.

The fixed point structure of the transformation in this example is very simple, but one can imagine that in more complicated problems the fixed point structure of various transformations might be highly non-trivial. This is certainly the case in statistical mechanics problems where renormalization methods are used to classify various phases of matters; the point here is that the renormalization group flow can be thought of as solving the pattern-recognition problem of classifying the long-range behavior of various statistical systems.

In summary, renormalization can be thought of as %is a special case of feature extraction and nearly sufficient statistics, treating small scales as noise, as such it is strictly 
a type of supervised learning\footnote{A subtlety regarding the above statements is presented by the Multi-scale Entanglement Renormalization Ansatz (MERA) \cite{mera}. MERA can be viewed as a variational class of wave functions whose parameters can be tuned to to match a given wave function as closely as possible. From this perspective, MERA is as an unsupervised machine learning algorithm, where classical probability distributions over many variables are replaced with quantum wavefunctions. Due to the special tensor network structure found in MERA, the resulting variational approximation of a given wavefunction has an interpretation as generating an RG flow. Hence this is an example of an unsupervised learning problem whose solution gives rise to an RG flow. This is only possible due to the extra mathematical structure in the problem (the specific tensor network found in MERA); a generic variational Ansatz does not give rise to any RG interpretation and vice versa.}, where the large scale properties of the system are considered the features. If the desired features are not large-scale properties (as in most machine learning cases), one might still expect the a generalized formalism of renormalization to provide some intuition to the problem by replacing a scale transformation with some other transformation. But calling some procedure renormalization or not is ultimately a matter of semantics; what remains to be seen is whether or not semantics has teeth, namely, whether the intuition about fixed points of the renormalization group flow can provide concrete insight into machine learning algorithms.
%there is no reason why existing renormalization schemes used in physics would provide any intuition about the problem. One might generalize the notion of renormalization to include not just transformations of scale but other more complicated examples.
%This makes it a special case of supervised learning, not unsupervised learning. We elaborate on this further in Appendix A, where we construct a counter-example to a recent claim \cite{mehta} that a so-called ``exact'' RG is equivalent to perfectly reconstructing the empirical probability distribution in an unsupervised problem.
%This makes it a special case of supervised learning, not unsupervised learning. %We elaborate on this further in Appendix A, where we address a recent suggestion \cite{mehta} that a so-called ``exact'' RG involves an unsupervised problem.
%The information-distillation nature of renormalization is explicit
In many numerical methods, the purpose of the renormalization group is to efficiently and accurately evaluate the free energy of the system as a function of macroscopic variables of interest such as temperature and pressure. Thus we can only sensibly talk about the accuracy of an RG-scheme once we have specified what macroscopic variables we are interested in. 
%In an unsupervised learning problem, there are no given macroscopic variables and hence we can only compare a single value of the free energy, which by itself has no physical meaning.

\subsection{No-flattening theorems}
\label{NoFlatteningSec}

Above we discussed how Markovian generative models cause $p(\wy|\ex)$ to be a composition of a number of simpler functions $f_i$.
Suppose that we can approximate each function $f_i$ with an efficient neural network for the reasons given in \Sec{ShallowSec}.
Then we can simply stack these networks on top of each other, to obtain an deep neural network efficiently approximating $p(\wy|\ex)$.

But is this the most efficient way to represent $p(\wy|\ex)$? Since we know that there are shallower networks that accurately approximate it, are any of these shallow networks as efficient as the deep one, or does flattening necessarily come at an efficiency cost?

To be precise, for a neural network $\f$ defined by \eq{NNeq}, we will say that the neural network $\f^\ell_\epsilon$ is the {\it flattened} version of $\f$ if 
its number $\ell$ of hidden layers is smaller and $\f^\ell_\epsilon$ approximates $\f$ within some error $\epsilon$ (as measured by some reasonable norm). We say that $\f^\ell_\epsilon$ is a {\it neuron-efficient flattening} if the sum of the dimensions of its hidden layers (sometimes referred to as the number of neurons $N_n$) is less than for $\f$. We say that $\f^\ell_\epsilon$ is a {\it synapse-efficient flattening} if the number $N_s$ of non-zero entries (sometimes called synapses) in its weight matrices is less than for $\f$. 
%We define the {\it flattening cost} as the ratio of the number of non-zero weights $C_w(l,\epsilon) = \min_{h_\epsilon^l} N_w(h_\epsilon^l)/N_w(h_\epsilon^l)$ or the ratio of number of neurons $C_n(l, \epsilon) = \min_{h_\epsilon^l} N_n(h_\epsilon^l)/N_n(h_\epsilon^l)$ of the optimal flattened neural network.
This lets us define the {\it flattening cost} of a network $\f$ as the two functions 
\beqa{FlatteningCostEq}
C_n(\f,\ell,\epsilon)\equiv\min_{\f_\epsilon^\ell} {N_n(\f_\epsilon^\ell)\over N_n(\f)},\\
C_s(\f,\ell,\epsilon)\equiv\min_{\f_\epsilon^\ell} {N_s(\f_\epsilon^\ell)\over N_s(\f)},
\eeqa
specifying the factor by which optimal flattening increases the neuron count and the synapse count, respectively.
We refer to results where $C_n>1$ or $C_s>1$ for some class of functions $\f$ as {\it``no-flattening theorems''}, since they imply that flattening comes at a cost and efficient flattening is impossible.
A complete list of no-flattening theorems would show exactly when deep networks are more efficient than shallow networks.

There has already been very interesting progress in this spirit, but crucial questions remain.
On one hand, it has been shown that deep is not always better, at least empirically for some image classification tasks \cite{reallydeep}. On the other hand, many functions $\f$ have been found for which the flattening cost is significant. 
Certain deep Boolean circuit networks are exponentially costly to flatten \cite{hastad1986almost}. 
Two families of multivariate polynomials with an exponential flattening cost $C_n$ are constructed in\cite{sumproduct}. \cite{poggio2015theory,poggio2016,mhaskarPoggio2016} focus on functions that have tree-like hierarchical compositional form, concluding that the flattening cost $C_n$ is exponential for almost all functions in Sobolev space. 
For the ReLU activation function, \cite{telgarsky2015representation} finds a class of functions that exhibit exponential flattening costs; \cite{montu14} study a tailored complexity measure of deep versus shallow ReLU networks.
\cite{eldan2015power} shows that given weak conditions on the activation function, there always exists at least one function that can be implemented in a 3-layer network which has an exponential flattening cost.
Finally, \cite{poole16, raghu16} study the differential geometry of shallow versus deep networks, and find 
that flattening is exponentially neuron-inefficient.
Further work elucidating the cost of flattening various classes of functions will clearly be highly valuable.

\subsection{Linear no-flattening theorems}

In the mean time, we will now see that interesting no-flattening results can be obtained even in the simpler-to-model context of {\it linear} neural networks \cite{saxe}, where the $\boldsymbol{\sigma}$ operators are replaced with the identity and all biases are set to zero such that $\A_i$ are simply linear operators (matrices). Every map is specified by a matrix of real (or complex) numbers, and composition is implemented by matrix multiplication. 

One might suspect that such a network is so simple that the questions concerning flattening become entirely trivial: after all,  successive multiplication with $n$ different matrices is equivalent to multiplying by a single matrix (their product).
%after all,  the composition of $n$ affine transformations is merely a single affine transformation $(\l=0)$, and successive multiplication with $n$ different matrices is equivalent to multiplying by a single matrix. 
While the effect of flattening is indeed trivial for {\it expressibility} ($\f$ can express any linear function, independently of how many layers there are), this is not the case for the {\it learnability}, which involves non-linear and complex dynamics despite the linearity of the network \cite{saxe}. We will show that the {\it efficiency} of such linear networks is also a very rich question.

Neuronal efficiency is trivially attainable for linear networks, since all hidden-layer neurons can be eliminated without accuracy loss by simply multiplying all the weight matrices together.
We will instead consider the case of synaptic efficiency and set $\ell=\epsilon=0$. 

Many divide-and-conquer algorithms in numerical linear algebra exploit some factorization of a particular matrix $\A$ in order to yield significant reduction in complexity. For example, when $\A$ represents the discrete Fourier transform (DFT), the fast Fourier transform (FFT) algorithm makes use of a sparse factorization of $\A$ which only contains $\mathcal{O}(n \log n)$ non-zero matrix elements instead of the naive single-layer implementation, which contains $n^2$ non-zero matrix elements. As first pointed out in \cite{bengio2007scaling}, this is an example where depth helps and, in our terminology, of a linear no-flattening theorem: fully flattening a network that performs an FFT of $n$ variables increases the synapse count $N_s$ 
from $\mathcal{O}(n \log n)$ to $\mathcal{O}(n^2)$, \ie, incurs a flattening cost 
$C_s=\mathcal{O}(n/\log n)\sim \mathcal{O}(n)$.
This argument applies also to many variants and generalizations of the FFT such as the Fast Wavelet Transform and the Fast Walsh-Hadamard Transform.

Another important example illustrating the subtlety of linear networks is matrix multiplication. More specifically, take the input of a neural network to be the entries of a matrix $\M$ and the output to be $\N \M$, where 
both $\M$ and $\N$ have size $n\times n$.
Since matrix multiplication is linear, this can be exactly implemented by a 1-layer linear neural network. Amazingly, the naive algorithm for matrix multiplication, which requires $n^3$ multiplications,
 is not optimal: the Strassen algorithm \cite{strassen1969gaussian} requires only $\mathcal{O}(n^\omega)$ multiplications (synapses), where $\omega = \log_2 7 \approx 2.81$, and recent work has cut 
this scaling exponent down to $\omega\approx 2.3728639$ \cite{le2014powers}.
This means that fully optimized matrix multiplication on a deep neural network has a flattening cost of at least
$C_s=\mathcal{O}(n^{0.6271361})$.

Low-rank matrix multiplication gives a more elementary no-flattening theorem. If $\A$ is a rank-$k$ matrix, we can factor it as $\A=\textbf{\B\C}$ where $\B$ is a $k \times n$ matrix and $\C$ is an $n \times k$ matrix. Hence the number of synapses is $n^2$ for an $\ell=0$ network and $2nk$ for an $\ell=1$-network, giving a flattening cost
$C_s=n/2k>1$ as long as the rank $k<n/2$.

Finally, let us consider flattening a network $\f=\A\B$, where $\A$ and $\B$ are random sparse $n\times n$ matrices such that each element is $1$ with probability $p$ and $0$ with probability $1-p$. Flattening the network results in a matrix 
$F_{ij} = \sum_k A_{ik} B_{kj}$, so the probability that $F_{ij}=0$ is $(1-p^2)^n$. Hence the number of non-zero components will on average be $\lp 1-(1-p^2)^n\rp n^2$, so
\beq{SparseFlatteningCostEq}
C_s = {\left[1-(1-p^2)^n\right]n^2\over 2n^2p}=
\frac{1-(1-p^2)^n}{2p}.
\eeq
Note that $C_s\le 1/2p$ and that this bound is asymptotically saturated for $n\gg 1/p^2$. Hence in the limit where $n$ is very large, flattening multiplication by sparse matrices $p \ll 1$ is horribly inefficient.

\subsection{A polynomial no-flattening theorem}

In \Sec{ShallowSec}, we saw that multiplication of two variables could be implemented by a flat neural network with 4 neurons in the hidden layer, using \eq{xyApproxEq} as illustrated in \fig{MultiplicationFig}.
In Appendix~\ref{PolynomialTheoremSec}, we show that \eq{xyApproxEq} is merely the $n=2$ special case of the formula
\beq{GeneralMultiplicationApproximatorEq}
\prod_{i=1}^n x_i = {1\over 2^n} \sum_{\{s\} } s_1...s_n\sigma(s_1 x_1 +...+s_n x_n),
\eeq
where the sum is over all possible $2^n$ configurations of $s_1, \cdots s_n$ where each $s_i$ can take on values $\pm 1$.
In other words, multiplication of $n$ variables can be implemented by a flat network with $2^n$ neurons in the hidden layer.
%We conjecture that this is construction is in fact optimal:
We also prove in Appendix~\ref{PolynomialTheoremSec} that this is the best one can do:
no neural network can implement an $n$-input multiplication gate using fewer than $2^n$ neurons in the hidden layer.
This is another powerful no-flattening theorem, telling us that polynomials are exponentially expensive to flatten. 
For example, if $n$ is a power of two, then the monomial $x_1 x_2...x_n$ can be evaluated by a deep network
using only $4n$ neurons arranged in a deep neural network where
$n$ copies of the multiplication gate from \fig{MultiplicationFig} are arranged in a binary tree with $\log_2 n$ layers (the 5th top neuron at the top of \fig{MultiplicationFig} need not be counted, as it is the input to whatever computation comes next).
In contrast, a functionally equivalent flattened network requires a whopping $2^n$ neurons. 
For example, a deep neural network can multiply 32 numbers using $4n=160$ neurons while a shallow one requires
$2^{32}=4,294,967,296$ neurons.
Since a broad class of real-world functions can be well approximated by polynomials, this helps explain why many useful neural networks cannot be efficiently flattened.

\section{Conclusions}
\label{ConclusionsSec}

We have shown that the success of deep and cheap (low-parameter-count) learning depends not only on mathematics but also on physics, which favors certain classes of exceptionally simple probability distributions that deep learning is uniquely suited to model. We argued that the success of shallow neural networks hinges on symmetry, locality,  and polynomial log-probability in data from or inspired by the natural world, which favors sparse low-order polynomial Hamiltonians that can be efficiently approximated. These arguments should be particularly relevant for explaining the success of machine-learning applications to physics, for example using a neural network to approximate a many-body wavefunction \cite{carleo}.
Whereas previous universality theorems guarantee that there exists a neural network that approximates any smooth function to within an error $\epsilon$, they cannot guarantee that the size of the neural network does not grow to infinity with shrinking $\epsilon$ or that the activation function $\sigma$ does not become pathological. We show constructively that given a multivariate polynomial and any generic non-linearity, a neural network with a {\it fixed} size and a generic smooth activation function can indeed approximate the polynomial highly efficiently.

%We then moved on to the question of cheap learning: efficiency. In order to answer the question of why cheap learning is possible, one must specify a generative model for the data, which we take to be a Markovian generative process. We argue that even if every step in the Markovian generative process is simple, the resulting data can be very complex. Future work should systematically study common steps in the generative process to quantify and tabulate exactly how complex certain steps that might be relevant for machine learning tasks are.

Turning to the separate question of depth, we have argued that the success of deep learning depends on the ubiquity of hierarchical and compositional generative processes in physics and other machine-learning applications. 
By studying the sufficient statistics of the generative process, we showed that the inference problem requires approximating a compositional function of the form $f_1 \circ f_2 \circ f_2 \circ \cdots$ that optimally distills out the information of interest from irrelevant noise in a hierarchical process that mirrors the generative process. 
Although such compositional functions can be efficiently implemented by a deep neural network as long as their individual steps can, it is generally {\it not} possible to retain the efficiency while flattening the network.
We extend existing ``no-flattening'' theorems \cite{sumproduct,poggio2016,mhaskarPoggio2016} by showing that efficient flattening is impossible even for many important cases involving {\it linear} networks. In particular, we prove that flattening polynomials is exponentially expensive, with $2^n$ neurons required to multiply $n$ numbers using a single hidden layer, a task that a deep network can perform using only $\sim 4n$ neurons.

Strengthening the analytic understanding of deep learning may suggest ways of improving it, both to make it more capable and to make it more robust. 
One promising area is to prove sharper and more comprehensive no-flattening theorems, placing lower and upper bounds on the cost of flattening networks implementing various classes of functions. 
%A concrete example is placing tight lower and upper bounds on the number of neurons and synaptic weights needed to approximate a given polynomial. 
%We conjecture that approximating a multiplication gate $x_1 x_2 \cdots x_n$ will require exponentially many neurons in $n$ using non-pathological activation functions, whereas we have shown that allowing for $\log_2 n$ layers allows us to use only $\sim 4 n$ neurons.

%
%* What we've done

%* Linear NNs are trivial in expressibility, but *not* in learnability (Ganguli) or efficiency (us)
 % (mention ganguli linear network paper for learnability)

%* dictionary table?

%Our work sharpens but leaves unanswered many questions. While we show that there exists a neural network with a fixed number of neurons that approximates a given polynomial arbitrarily well.

%* Outlook: open questions. The unsolved core of the problem. 
  %- prove that all the common ML-steps can be efficiently implemented
 
   % - probe no-flattening theorems
 
    %    - conjecture: need $2^n$ neurons to nail $x_1... x_n$
 
     %   - weaker conjecture: need $>4n$ to nail $x_1... x_n$
 
   %- everything about learning

\bigskip

%\section{Acknowledgements}
{\bf Acknowledgements:}
This work was supported by the Foundational Questions Institute \url{http://fqxi.org/}, the Rothberg Family Fund for Cognitive Science and NSF grant 1122374. We thank Scott Aaronson, Frank Ban, Yoshua Bengio, Rico Jonschkowski, Tomaso Poggio, Bart Selman,  Viktoriya Krakovna, Krishanu Sankar and Boya Song for helpful discussions and suggestions, Frank Ban, Fernando Perez, Jared Jolton, and the anonymous referee for helpful corrections and the Center for Brains, Minds, and Machines (CBMM) for hospitality.

\appendix

\section{The polynomial no-flattening theorem}
\label{PolynomialTheoremSec}

We saw above that a neural network can compute polynomials accurately and efficiently at linear cost, using only about 4 neurons per multiplication. For example, if $n$ is a power of two, then the monomial $\prod_{i=1}^n x_i$ can be evaluated using $4n$ neurons arranged in a binary tree network with $\log_2 n$ hidden layers. In this appendix, we will prove a no-flattening theorem demonstrating that flattening polynomials is exponentially expensive:

{\bf Theorem}: Suppose we are using a generic smooth activation function $\sigma(x) = \sum_{k=0}^\infty \sigma_k x^k$, where $\sigma_k \ne 0$ for $0\le k\le n$. Then for any desired accuracy $\epsilon>0$, there exists a neural network that can implement the function $\prod_{i=1}^n x_i$ using a single hidden layer of $2^n$ neurons.  Furthermore, this is the smallest possible number of neurons in any such network with only a single hidden layer.

This result may be compared to problems in Boolean circuit complexity, notably the question of whether $TC^0 = TC^1$ \cite{vollmer2013introduction}. Here circuit depth is analogous to number of layers, and the number of gates is analogous to the number of neurons. In both the Boolean circuit model and the neural network model, one is allowed to use neurons/gates which have an unlimited number of inputs. The constraint in the definition of $TC^i$ that each of the gate elements be from a standard universal library (\texttt{AND}, \texttt{OR}, \texttt{NOT}, \texttt{Majority}) is analogous to our constraint to use a particular nonlinear function. Note, however, that our theorem is weaker by applying only to depth 1, while $TC^0$ includes all circuits of depth $O(1)$.

% This theorem might be viewed as the neural-network analog of the Boolean circuit statement that the complexity class TC$^0$ is a strict subset of TC$^1$\footnote{We thank Scott Aaronson for bringing this to our attention.}. More explicitly, this theorem implies that there exists problems which can be solved by neural networks with only a polynomial number of neurons for depth $O(\log N)$, but cannot be solved with a polynomial number of neurons for an $O(1)$ depth. Here, circuit depth is analogous to number of layers, and the number of gates is analogous the number of neurons. In both the Boolean circuit model and neural network model, we are allowed to use neurons/gates which have an unlimited number of inputs. The constraint in the definition of $TC^i$ that each of the gate elements be one from a standard universal library (AND, OR, NOT, MAJORITY) is analogous to our constraint that one only uses a particular sigmoid function. Although this analogy is quite close morally speaking, there is a key difference: remarkably, separating this polynomial hierarchy is apparently much easier in the neural complexity case than in the Boolean circuit case.

%that a flattened network can compute $\prod_{i=1}^n x_i$ using $2^n$ neurons in the hidden layer, but no fewer.

%%%%%%%%%%%%%%%%%%%%%%%%%

\subsection{Proof that $2^n$ neurons are sufficient}

A neural network with a single hidden layer of $m$ neurons that approximates a product gate for $n$ inputs can be formally written as a choice of constants $a_{ij}$ and $w_j$ satisfying
\begin{equation}
\sum_{j=1}^m w_j \sigma\left(\sum_{i=1}^n a_{ij} x_i\right) \approx \prod_{i=1}^n x_i. \label{goal}
\end{equation}
Here, we use $\approx$ to denote that the two sides of (\ref{goal}) have identical Taylor expansions up to terms of degree $n$; as we discussed earlier in our construction of a product gate for two inputs, this exables us to achieve arbitrary accuracy $\epsilon$ by first scaling down the factors $x_i$, then approximately multiplying them and finally scaling up the result.

We may expand (\ref{goal}) using the definition $\sigma(x) = \sum_{k=0}^\infty \sigma_k x^k$ and drop terms of the Taylor expansion with degree greater than $n$, since they do not affect the approximation.  Thus, we wish to find the minimal $m$ such that there exist constants $a_{ij}$ and $w_j$ satisfying
\begin{eqnarray}
\sigma_n\sum_{j=1}^m w_j\left(\sum_{i=1}^n a_{ij} x_i\right)^n &=& \prod_{i=1}^n x_i, \label{goaln} \\
\sigma_k\sum_{j=1}^m w_j\left(\sum_{i=1}^n a_{ij} x_i\right)^k &=& 0,\label{goalk} 
\end{eqnarray}
for all  $0\le k \le n-1$. 
Let us set $m = 2^n$, and enumerate the subsets of $\{1,\ldots,n\}$ as $S_1,\ldots,S_m$ in some order.  Define a network of $m$ neurons in a single hidden layer by setting $a_{ij}$ equal to the function $s_i(S_j)$ which is 
$-1$ if $i\in S_j$ and $+1$ otherwise, setting 
%$w_j = (-1)^{|S_j|} / (2^n \cdot n!\cdot \sigma_n)$
\begin{equation}
w_j\equiv {1\over 2^n n!\sigma_n}\prod_{i=1}^n a_{ij}
=
{(-1)^{|S_j|}\over 2^n n!\sigma_n}.
\end{equation}
In other words, up to an overall normalization constant, all coefficients $a_{ij}$ and $w_j$ equal $\pm 1$, and each weight $w_j$ is simply the product of the corresponding $a_{ij}$.

We must prove that this network indeed satisfies  equations (\ref{goaln}) and (\ref{goalk}). 
The essence of our proof will be to expand the left hand side of Equation~(\ref{goal}) and show that all monomial terms except $x_1\cdot\cdot\cdot x_n$ come in pairs that cancel.
To show this, consider a single monomial $p(\textbf{x}) = x_1^{r_1}\cdots x_n^{r_n}$ where $r_1 + \ldots + r_n = r \le n$.

If $p(\textbf{x}) \ne \prod_{i=1}^n x_i$, then we must show that the coefficient of $p(\textbf{x})$ in $\sigma_r\sum_{j=1}^m w_j\left(\sum_{i=1}^n a_{ij} x_i\right)^r$ is 0. Since $p(\textbf{x}) \ne \prod_{i=1}^n x_i$, there must be some $i_0$ such that $r_{i_0} = 0$. In other words, $p(\textbf{x})$ does not depend on the variable $x_{i_0}$.
Since the sum in Equation~(\ref{goal}) is over all combinations of $\pm$ signs for all variables, every term will be canceled by another term where the (non-present) $x_{i_0}$ has the opposite sign and the weight $w_j$ has the opposite sign: 
%Specifically,we can manipulate the expression in question as shown in 

% (\ref{multicol}).
% \begin{table*}[htbp]
% \begin{eqnarray}
% \sigma_r\sum_{j=1}^m w_j\left(\sum_{i=1}^n a_{ij} x_i\right)^r 
% &=& \sigma_r\sum_{S_j} {(-1)^{|S_j|}\over 2^n n!\sigma_r}\left(\sum_{i=1}^n s_i(S_j) x_i\right)^r \nonumber\\
% &=& \sigma_r\sum_{S_j\not\ni i_0} \left[{(-1)^{|S_j|}\over 2^n n!\sigma_r}\left(\sum_{i=1}^n s_i(S_j) x_i\right)^r
% + {(-1)^{|S_j \cup \{i_0\}|}\over 2^n n!\sigma_r}\left(\sum_{i=1}^n s_i(S_j \cup \{i_0\}) x_i\right)^r\right] \nonumber\\
% &=& \sum_{S_j\not\ni i_0} {(-1)^{|S_j|}\over 2^n n!}\left[\left(\sum_{i=1}^n s_i(S_j) x_i\right)^r -\left(\sum_{i=1}^n s_i(S_j \cup \{i_0\}) x_i\right)^r\right]\label{multicol}
% \end{eqnarray}
% \end{table*}

\eqn{\sigma_r\sum_{j=1}^m w_j & \left(\sum_{i=1}^n a_{ij} x_i\right)^r \\
&= \sigma_r\sum_{S_j} {(-1)^{|S_j|}\over 2^n n!\sigma_r}\left(\sum_{i=1}^n s_i(S_j) x_i\right)^r \nonumber\\
&= \sigma_r\sum_{S_j\not\ni i_0} \Bigg[{(-1)^{|S_j|}\over 2^n n!\sigma_r}\left(\sum_{i=1}^n s_i(S_j) x_i\right)^r\\
&+ {(-1)^{|S_j \cup \{i_0\}|}\over 2^n n!\sigma_r}\left(\sum_{i=1}^n s_i(S_j \cup \{i_0\}) x_i\right)^r\Bigg] \nonumber\\
&= \sum_{S_j\not\ni i_0} {(-1)^{|S_j|}\over 2^n n!}\Bigg[\left(\sum_{i=1}^n s_i(S_j) x_i\right)^r \\
&-\left(\sum_{i=1}^n s_i(S_j \cup \{i_0\}) x_i\right)^r\Bigg]
\label{multicol}
}

Observe that the coefficient of $p(\textbf{x})$ is equal in $\left(\sum_{i=1}^n s_i(S_j) x_i\right)^r$ and $\left(\sum_{i=1}^n s_i(S_j \cup \{i_0\}) x_i\right)^r$, since $r_{i_0}=0$. Therefore, the overall coefficient of $p(\textbf{x})$ in the above expression must vanish, which implies that (\ref{goalk}) is satisfied.

If instead $p(\textbf{x}) = \prod_{i=1}^n x_i$, then all terms have the coefficient of $p(\textbf{x})$ in $\left(\sum_{i=1}^n a_{ij} x_i\right)^n$ is $n!\, \prod_{i=1}^n a_{ij} = (-1)^{|S_j|} n!$, because all $n!$ terms are identical and there is no cancelation.  Hence, the coefficient of $p(\textbf{x})$ on the left-hand side of (\ref{goaln}) is
$$
\sigma_n\sum_{j=1}^m \frac{(-1)^{|S_j|}}{2^n n! \sigma_n} (-1)^{|S_j|} n! = 1,
$$
completing our proof that this network indeed approximates the desired product gate.

From the standpoint of group theory, our construction involves a representation of the group $G = \mathbb{Z}_2^n$, acting upon the space of polynomials in the variables $x_1, x_2, \ldots, x_n$.  The group $G$ is generated by elements $g_i$ such that $g_i$ flips the sign of $x_i$ wherever it occurs.  Then, our construction corresponds to the computation
$$
\textbf{f}(x_1,\ldots,x_n) = (1 - g_1)(1 - g_2) \cdots (1-g_n)\sigma(x_1+x_2+\ldots+x_n).
$$
Every monomial of degree at most $n$, with the exception of the product $x_1\cdots x_n$, is sent to 0 by $(1 - g_i)$ for at least one choice of $i$. Therefore, $\textbf{f}(x_1,\ldots,x_n)$ approximates a product gate (up to a normalizing constant).

%%%%%%%%%%%%%%%%%%%%%%%%%

\subsection{Proof that $2^n$ neurons are necessary}

Suppose that $S$ is a subset of $\{1,\ldots,n\}$ and consider taking the partial derivatives of (\ref{goaln}) and (\ref{goalk}), respectively, with respect to all the variables $\{x_h\}_{h\in S}$.  Then, we obtain
{\small
\begin{eqnarray}
\frac{n!\, \sigma_n}{|n-S|!} \sum_{j=1}^m w_j\prod_{h\in S} a_{hj}\left(\sum_{i=1}^n a_{ij} x_i\right)^{n-|S|} &=& \prod_{h\not\in S} x_h, \label{derivn}\\
\frac{k!\, \sigma_k}{|k-S|!} \sum_{j=1}^m w_j\prod_{h\in S} a_{hj}\left(\sum_{i=1}^n a_{ij} x_i\right)^{k-|S|} &=& 0, \label{derivk}
\end{eqnarray}}for all $0\le k \le n-1$. Let $\A$ denote the $2^n \times m$ matrix with elements \begin{equation}
\label{AdefEq}
A_{Sj} \equiv \prod_{h \in S} a_{hj}.
\end{equation}
We will show that $\A$ has full row rank.  Suppose, towards contradiction, that
$\c^t\A={\bf 0}$ for some non-zero vector $\c$. Specifically, suppose that there is a linear dependence between rows of $\A$ given by
\begin{equation}
\label{cEq}
\sum_{\ell=1}^r c_{\ell} A_{S_\ell,j} = 0,
\end{equation}
where the $S_{\ell}$ are distinct and $c_{\ell} \ne 0$ for every $\ell$.  Let $s$ be the maximal cardinality of any $S_{\ell}$. 
Defining the vector $\d$ whose components are
\begin{equation}
d_j\equiv w_j\left(\sum_{i=1}^n a_{ij} x_i\right)^{n-s},
\end{equation}
taking the dot product of \eq{cEq} with $\d$
gives
%Consider taking the inner product of each side of the above equality by the vector with entries (indexed by $j$) equal to $w_j\left(\sum_{i=1}^n a_{ij} x_i\right)^{n-s}$:
\begin{eqnarray}
\label{ProofEq1}
0&=&\c^t\A\d = \sum_{\ell=1}^r c_{\ell} \sum_{j=1}^m w_j\prod_{h \in S_{\ell}} a_{hj} \left(\sum_{i=1}^n a_{ij} x_i\right)^{n-s} \nonumber\\
&=& \sum_{\ell \mid (|S_{\ell}| = s)} c_{\ell} \sum_{j=1}^m w_j\prod_{h\in S_{\ell}} a_{hj} \left(\sum_{i=1}^n a_{ij} x_i\right)^{n-|S_{\ell}|} \\
&& + \sum_{\ell \mid (|S_{\ell}| < s)} c_{\ell} \sum_{j=1}^m w_j\prod_{h\in S_{\ell}} a_{hj} \left(\sum_{i=1}^n a_{ij} x_i\right)^{(n+|S_{\ell}|-s)-|S_{\ell}|}.\nonumber
\end{eqnarray}
%Applying (\ref{derivn}) to the first term and (\ref{derivk}) (with $k=n+|S_{\ell}|-s$) to the second term, we obtain:
Applying \eq{derivk} (with $k=n+|S_{\ell}|-s$) shows that the second term vanishes.
Substituting \eq{derivn} now simplifies \eq{ProofEq1} to 
\begin{equation}
0 = \sum_{\ell \mid (|S_{\ell}| = s)} \frac{c_{\ell}|n-S_{\ell}|!}{n!\> \sigma_n} \> \prod_{h\not\in S_{\ell}} x_h,
\end{equation}
\ie, to a statement that a set of monomials are linearly dependent.
Since all distinct monomials are in fact linearly independent, this
is a contradiction of our assumption that the $S_{\ell}$ are distinct and $c_{\ell}$ are nonzero.  We conclude that $\A$ has full row rank, and therefore that $m \ge 2^n$, which concludes the proof.

\bibliography{cheap}

\begin{thebibliography}{50}
\expandafter\ifx\csname natexlab\endcsname\relax\def\natexlab#1{#1}\fi
\expandafter\ifx\csname bibnamefont\endcsname\relax
  \def\bibnamefont#1{#1}\fi
\expandafter\ifx\csname bibfnamefont\endcsname\relax
  \def\bibfnamefont#1{#1}\fi
\expandafter\ifx\csname citenamefont\endcsname\relax
  \def\citenamefont#1{#1}\fi
\expandafter\ifx\csname url\endcsname\relax
  \def\url#1{\texttt{#1}}\fi
\expandafter\ifx\csname urlprefix\endcsname\relax\def\urlprefix{URL }\fi
\providecommand{\bibinfo}[2]{#2}
\providecommand{\eprint}[2][]{\url{#2}}

\bibitem[{\citenamefont{LeCun et~al.}(2015)\citenamefont{LeCun, Bengio, and
  Hinton}}]{lecun2015deep}
\bibinfo{author}{\bibfnamefont{Y.}~\bibnamefont{LeCun}},
  \bibinfo{author}{\bibfnamefont{Y.}~\bibnamefont{Bengio}}, \bibnamefont{and}
  \bibinfo{author}{\bibfnamefont{G.}~\bibnamefont{Hinton}},
  \bibinfo{journal}{Nature} \textbf{\bibinfo{volume}{521}},
  \bibinfo{pages}{436} (\bibinfo{year}{2015}).

\bibitem[{\citenamefont{Bengio}(2009)}]{bengio2009}
\bibinfo{author}{\bibfnamefont{Y.}~\bibnamefont{Bengio}},
  \bibinfo{journal}{Foundations and trends{\textregistered} in Machine
  Learning} \textbf{\bibinfo{volume}{2}}, \bibinfo{pages}{1}
  (\bibinfo{year}{2009}).

\bibitem[{\citenamefont{Russell et~al.}(2015)\citenamefont{Russell, Dewey, and
  Tegmark}}]{russell2015research}
\bibinfo{author}{\bibfnamefont{S.}~\bibnamefont{Russell}},
  \bibinfo{author}{\bibfnamefont{D.}~\bibnamefont{Dewey}}, \bibnamefont{and}
  \bibinfo{author}{\bibfnamefont{M.}~\bibnamefont{Tegmark}},
  \bibinfo{journal}{AI Magazine} \textbf{\bibinfo{volume}{36}}
  (\bibinfo{year}{2015}).

\bibitem[{\citenamefont{Herbrich and Williamson}(2002)}]{luck}
\bibinfo{author}{\bibfnamefont{R.}~\bibnamefont{Herbrich}} \bibnamefont{and}
  \bibinfo{author}{\bibfnamefont{R.~C.} \bibnamefont{Williamson}},
  \bibinfo{journal}{Journal of Machine Learning Research}
  \textbf{\bibinfo{volume}{3}}, \bibinfo{pages}{175} (\bibinfo{year}{2002}).

\bibitem[{\citenamefont{Shawe-Taylor et~al.}(1998)\citenamefont{Shawe-Taylor,
  Bartlett, Williamson, and Anthony}}]{shawe1998}
\bibinfo{author}{\bibfnamefont{J.}~\bibnamefont{Shawe-Taylor}},
  \bibinfo{author}{\bibfnamefont{P.~L.} \bibnamefont{Bartlett}},
  \bibinfo{author}{\bibfnamefont{R.~C.} \bibnamefont{Williamson}},
  \bibnamefont{and} \bibinfo{author}{\bibfnamefont{M.}~\bibnamefont{Anthony}},
  \bibinfo{journal}{IEEE transactions on Information Theory}
  \textbf{\bibinfo{volume}{44}}, \bibinfo{pages}{1926} (\bibinfo{year}{1998}).

\bibitem[{\citenamefont{Poggio et~al.}(2015)\citenamefont{Poggio, Anselmi, and
  Rosasco}}]{poggio2015theory}
\bibinfo{author}{\bibfnamefont{T.}~\bibnamefont{Poggio}},
  \bibinfo{author}{\bibfnamefont{F.}~\bibnamefont{Anselmi}}, \bibnamefont{and}
  \bibinfo{author}{\bibfnamefont{L.}~\bibnamefont{Rosasco}},
  \bibinfo{type}{Tech. Rep.}, \bibinfo{institution}{Center for Brains, Minds
  and Machines (CBMM)} (\bibinfo{year}{2015}).

\bibitem[{\citenamefont{{Mehta} and {Schwab}}(2014)}]{mehta}
\bibinfo{author}{\bibfnamefont{P.}~\bibnamefont{{Mehta}}} \bibnamefont{and}
  \bibinfo{author}{\bibfnamefont{D.~J.} \bibnamefont{{Schwab}}},
  \bibinfo{journal}{ArXiv e-prints}  (\bibinfo{year}{2014}),
  \eprint{1410.3831}.

\bibitem[{\citenamefont{Hornik et~al.}(1989)\citenamefont{Hornik, Stinchcombe,
  and White}}]{hornik1989multilayer}
\bibinfo{author}{\bibfnamefont{K.}~\bibnamefont{Hornik}},
  \bibinfo{author}{\bibfnamefont{M.}~\bibnamefont{Stinchcombe}},
  \bibnamefont{and} \bibinfo{author}{\bibfnamefont{H.}~\bibnamefont{White}},
  \bibinfo{journal}{Neural networks} \textbf{\bibinfo{volume}{2}},
  \bibinfo{pages}{359} (\bibinfo{year}{1989}).

\bibitem[{\citenamefont{Cybenko}(1989)}]{cybenko1989approximation}
\bibinfo{author}{\bibfnamefont{G.}~\bibnamefont{Cybenko}},
  \bibinfo{journal}{Mathematics of control, signals and systems}
  \textbf{\bibinfo{volume}{2}}, \bibinfo{pages}{303} (\bibinfo{year}{1989}).

\bibitem[{\citenamefont{Pinkus}(1999)}]{pinkus1999approximation}
\bibinfo{author}{\bibfnamefont{A.}~\bibnamefont{Pinkus}},
  \bibinfo{journal}{Acta Numerica} \textbf{\bibinfo{volume}{8}},
  \bibinfo{pages}{143} (\bibinfo{year}{1999}).

\bibitem[{\citenamefont{Gnedenko et~al.}(1954)\citenamefont{Gnedenko,
  Kolmogorov, Gnedenko, and Kolmogorov}}]{clt}
\bibinfo{author}{\bibfnamefont{B.}~\bibnamefont{Gnedenko}},
  \bibinfo{author}{\bibfnamefont{A.}~\bibnamefont{Kolmogorov}},
  \bibinfo{author}{\bibfnamefont{B.}~\bibnamefont{Gnedenko}}, \bibnamefont{and}
  \bibinfo{author}{\bibfnamefont{A.}~\bibnamefont{Kolmogorov}},
  \bibinfo{journal}{Amer. J. Math.} \textbf{\bibinfo{volume}{105}},
  \bibinfo{pages}{28} (\bibinfo{year}{1954}).

\bibitem[{\citenamefont{Jaynes}(1957)}]{Jaynes}
\bibinfo{author}{\bibfnamefont{E.~T.} \bibnamefont{Jaynes}},
  \bibinfo{journal}{Physical review} \textbf{\bibinfo{volume}{106}},
  \bibinfo{pages}{620} (\bibinfo{year}{1957}).

\bibitem[{\citenamefont{Tegmark et~al.}(2006)\citenamefont{Tegmark, Aguirre,
  Rees, and Wilczek}}]{axions}
\bibinfo{author}{\bibfnamefont{M.}~\bibnamefont{Tegmark}},
  \bibinfo{author}{\bibfnamefont{A.}~\bibnamefont{Aguirre}},
  \bibinfo{author}{\bibfnamefont{M.~J.} \bibnamefont{Rees}}, \bibnamefont{and}
  \bibinfo{author}{\bibfnamefont{F.}~\bibnamefont{Wilczek}},
  \bibinfo{journal}{Physical Review D} \textbf{\bibinfo{volume}{73}},
  \bibinfo{pages}{023505} (\bibinfo{year}{2006}).

\bibitem[{\citenamefont{Delalleau and Bengio}(2011)}]{sumproduct}
\bibinfo{author}{\bibfnamefont{O.}~\bibnamefont{Delalleau}} \bibnamefont{and}
  \bibinfo{author}{\bibfnamefont{Y.}~\bibnamefont{Bengio}}, in
  \emph{\bibinfo{booktitle}{Advances in Neural Information Processing Systems}}
  (\bibinfo{year}{2011}), pp. \bibinfo{pages}{666--674}.

\bibitem[{\citenamefont{{Mhaskar} et~al.}(2016)\citenamefont{{Mhaskar}, {Liao},
  and {Poggio}}}]{poggio2016}
\bibinfo{author}{\bibfnamefont{H.}~\bibnamefont{{Mhaskar}}},
  \bibinfo{author}{\bibfnamefont{Q.}~\bibnamefont{{Liao}}}, \bibnamefont{and}
  \bibinfo{author}{\bibfnamefont{T.}~\bibnamefont{{Poggio}}},
  \bibinfo{journal}{ArXiv e-prints}  (\bibinfo{year}{2016}),
  \eprint{1603.00988}.

\bibitem[{\citenamefont{Mhaskar and Poggio}(2016)}]{mhaskarPoggio2016}
\bibinfo{author}{\bibfnamefont{H.}~\bibnamefont{Mhaskar}} \bibnamefont{and}
  \bibinfo{author}{\bibfnamefont{T.}~\bibnamefont{Poggio}},
  \bibinfo{journal}{arXiv preprint arXiv:1608.03287}  (\bibinfo{year}{2016}).

\bibitem[{\citenamefont{Adam et~al.}(2015)\citenamefont{Adam, Ade, Aghanim,
  Akrami, Alves, Arnaud, Arroja, Aumont, Baccigalupi, Ballardini
  et~al.}}]{Planck2015Results}
\bibinfo{author}{\bibfnamefont{R.}~\bibnamefont{Adam}},
  \bibinfo{author}{\bibfnamefont{P.}~\bibnamefont{Ade}},
  \bibinfo{author}{\bibfnamefont{N.}~\bibnamefont{Aghanim}},
  \bibinfo{author}{\bibfnamefont{Y.}~\bibnamefont{Akrami}},
  \bibinfo{author}{\bibfnamefont{M.}~\bibnamefont{Alves}},
  \bibinfo{author}{\bibfnamefont{M.}~\bibnamefont{Arnaud}},
  \bibinfo{author}{\bibfnamefont{F.}~\bibnamefont{Arroja}},
  \bibinfo{author}{\bibfnamefont{J.}~\bibnamefont{Aumont}},
  \bibinfo{author}{\bibfnamefont{C.}~\bibnamefont{Baccigalupi}},
  \bibinfo{author}{\bibfnamefont{M.}~\bibnamefont{Ballardini}},
  \bibnamefont{et~al.}, \bibinfo{journal}{arXiv preprint arXiv:1502.01582}
  (\bibinfo{year}{2015}).

\bibitem[{\citenamefont{Seljak and Zaldarriaga}(1996)}]{cmbfast}
\bibinfo{author}{\bibfnamefont{U.}~\bibnamefont{Seljak}} \bibnamefont{and}
  \bibinfo{author}{\bibfnamefont{M.}~\bibnamefont{Zaldarriaga}},
  \bibinfo{journal}{arXiv preprint astro-ph/9603033}  (\bibinfo{year}{1996}).

\bibitem[{\citenamefont{Tegmark}(1997{\natexlab{a}})}]{cl}
\bibinfo{author}{\bibfnamefont{M.}~\bibnamefont{Tegmark}},
  \bibinfo{journal}{Physical Review D} \textbf{\bibinfo{volume}{55}},
  \bibinfo{pages}{5895} (\bibinfo{year}{1997}{\natexlab{a}}).

\bibitem[{\citenamefont{Bond et~al.}(1998)\citenamefont{Bond, Jaffe, and
  Knox}}]{BJK}
\bibinfo{author}{\bibfnamefont{J.}~\bibnamefont{Bond}},
  \bibinfo{author}{\bibfnamefont{A.~H.} \bibnamefont{Jaffe}}, \bibnamefont{and}
  \bibinfo{author}{\bibfnamefont{L.}~\bibnamefont{Knox}},
  \bibinfo{journal}{Physical Review D} \textbf{\bibinfo{volume}{57}},
  \bibinfo{pages}{2117} (\bibinfo{year}{1998}).

\bibitem[{\citenamefont{Tegmark et~al.}(2003)\citenamefont{Tegmark,
  de~Oliveira-Costa, and Hamilton}}]{mapforegs}
\bibinfo{author}{\bibfnamefont{M.}~\bibnamefont{Tegmark}},
  \bibinfo{author}{\bibfnamefont{A.}~\bibnamefont{de~Oliveira-Costa}},
  \bibnamefont{and} \bibinfo{author}{\bibfnamefont{A.~J.}
  \bibnamefont{Hamilton}}, \bibinfo{journal}{Physical Review D}
  \textbf{\bibinfo{volume}{68}}, \bibinfo{pages}{123523}
  (\bibinfo{year}{2003}).

\bibitem[{\citenamefont{Ade et~al.}(2014)\citenamefont{Ade, Aghanim,
  Armitage-Caplan, Arnaud, Ashdown, Atrio-Barandela, Aumont, Baccigalupi,
  Banday, Barreiro et~al.}}]{PlanckForegs}
\bibinfo{author}{\bibfnamefont{P.}~\bibnamefont{Ade}},
  \bibinfo{author}{\bibfnamefont{N.}~\bibnamefont{Aghanim}},
  \bibinfo{author}{\bibfnamefont{C.}~\bibnamefont{Armitage-Caplan}},
  \bibinfo{author}{\bibfnamefont{M.}~\bibnamefont{Arnaud}},
  \bibinfo{author}{\bibfnamefont{M.}~\bibnamefont{Ashdown}},
  \bibinfo{author}{\bibfnamefont{F.}~\bibnamefont{Atrio-Barandela}},
  \bibinfo{author}{\bibfnamefont{J.}~\bibnamefont{Aumont}},
  \bibinfo{author}{\bibfnamefont{C.}~\bibnamefont{Baccigalupi}},
  \bibinfo{author}{\bibfnamefont{A.~J.} \bibnamefont{Banday}},
  \bibinfo{author}{\bibfnamefont{R.}~\bibnamefont{Barreiro}},
  \bibnamefont{et~al.}, \bibinfo{journal}{Astronomy \& Astrophysics}
  \textbf{\bibinfo{volume}{571}}, \bibinfo{pages}{A12} (\bibinfo{year}{2014}).

\bibitem[{\citenamefont{Tegmark}(1997{\natexlab{b}})}]{mapmaking}
\bibinfo{author}{\bibfnamefont{M.}~\bibnamefont{Tegmark}},
  \bibinfo{journal}{The Astrophysical Journal Letters}
  \textbf{\bibinfo{volume}{480}}, \bibinfo{pages}{L87}
  (\bibinfo{year}{1997}{\natexlab{b}}).

\bibitem[{\citenamefont{Hinshaw et~al.}(2003)\citenamefont{Hinshaw, Barnes,
  Bennett, Greason, Halpern, Hill, Jarosik, Kogut, Limon, Meyer
  et~al.}}]{WMAPmapmaking}
\bibinfo{author}{\bibfnamefont{G.}~\bibnamefont{Hinshaw}},
  \bibinfo{author}{\bibfnamefont{C.}~\bibnamefont{Barnes}},
  \bibinfo{author}{\bibfnamefont{C.}~\bibnamefont{Bennett}},
  \bibinfo{author}{\bibfnamefont{M.}~\bibnamefont{Greason}},
  \bibinfo{author}{\bibfnamefont{M.}~\bibnamefont{Halpern}},
  \bibinfo{author}{\bibfnamefont{R.}~\bibnamefont{Hill}},
  \bibinfo{author}{\bibfnamefont{N.}~\bibnamefont{Jarosik}},
  \bibinfo{author}{\bibfnamefont{A.}~\bibnamefont{Kogut}},
  \bibinfo{author}{\bibfnamefont{M.}~\bibnamefont{Limon}},
  \bibinfo{author}{\bibfnamefont{S.}~\bibnamefont{Meyer}},
  \bibnamefont{et~al.}, \bibinfo{journal}{The Astrophysical Journal Supplement
  Series} \textbf{\bibinfo{volume}{148}}, \bibinfo{pages}{63}
  (\bibinfo{year}{2003}).

\bibitem[{\citenamefont{Hinton}(2010)}]{hinton2010practical}
\bibinfo{author}{\bibfnamefont{G.}~\bibnamefont{Hinton}},
  \bibinfo{journal}{Momentum} \textbf{\bibinfo{volume}{9}},
  \bibinfo{pages}{926} (\bibinfo{year}{2010}).

\bibitem[{\citenamefont{{\'E}mile~Borel}(1909)}]{Borel1909}
\bibinfo{author}{\bibfnamefont{M.}~\bibnamefont{{\'E}mile~Borel}},
  \bibinfo{journal}{Rendiconti del Circolo Matematico di Palermo (1884-1940)}
  \textbf{\bibinfo{volume}{27}}, \bibinfo{pages}{247} (\bibinfo{year}{1909}).

\bibitem[{\citenamefont{Fisher}(1922)}]{fisher}
\bibinfo{author}{\bibfnamefont{R.~A.} \bibnamefont{Fisher}},
  \bibinfo{journal}{Philosophical Transactions of the Royal Society of London.
  Series A, Containing Papers of a Mathematical or Physical Character}
  \textbf{\bibinfo{volume}{222}}, \bibinfo{pages}{309} (\bibinfo{year}{1922}).

\bibitem[{\citenamefont{Riesenhuber and Poggio}(2000)}]{riesenhuber2000models}
\bibinfo{author}{\bibfnamefont{M.}~\bibnamefont{Riesenhuber}} \bibnamefont{and}
  \bibinfo{author}{\bibfnamefont{T.}~\bibnamefont{Poggio}},
  \bibinfo{journal}{Nature neuroscience} \textbf{\bibinfo{volume}{3}},
  \bibinfo{pages}{1199} (\bibinfo{year}{2000}).

\bibitem[{\citenamefont{Kullback and Leibler}(1951)}]{kldiv}
\bibinfo{author}{\bibfnamefont{S.}~\bibnamefont{Kullback}} \bibnamefont{and}
  \bibinfo{author}{\bibfnamefont{R.~A.} \bibnamefont{Leibler}},
  \bibinfo{journal}{Ann. Math. Statist.} \textbf{\bibinfo{volume}{22}},
  \bibinfo{pages}{79} (\bibinfo{year}{1951}),
  \urlprefix\url{http://dx.doi.org/10.1214/aoms/1177729694}.

\bibitem[{\citenamefont{Cover and Thomas}(2012)}]{coverthomas}
\bibinfo{author}{\bibfnamefont{T.~M.} \bibnamefont{Cover}} \bibnamefont{and}
  \bibinfo{author}{\bibfnamefont{J.~A.} \bibnamefont{Thomas}},
  \emph{\bibinfo{title}{Elements of information theory}}
  (\bibinfo{publisher}{John Wiley \&amp; Sons}, \bibinfo{year}{2012}).

\bibitem[{\citenamefont{Kardar}(2007)}]{kardar}
\bibinfo{author}{\bibfnamefont{M.}~\bibnamefont{Kardar}},
  \emph{\bibinfo{title}{Statistical physics of fields}}
  (\bibinfo{publisher}{Cambridge University Press}, \bibinfo{year}{2007}).

\bibitem[{\citenamefont{Cardy}(1996)}]{cardy1996}
\bibinfo{author}{\bibfnamefont{J.}~\bibnamefont{Cardy}},
  \emph{\bibinfo{title}{Scaling and renormalization in statistical physics}},
  vol.~\bibinfo{volume}{5} (\bibinfo{publisher}{Cambridge university press},
  \bibinfo{year}{1996}).

\bibitem[{\citenamefont{{Johnson} et~al.}(2007)\citenamefont{{Johnson},
  {Malioutov}, and {Willsky}}}]{johnson07}
\bibinfo{author}{\bibfnamefont{J.~K.} \bibnamefont{{Johnson}}},
  \bibinfo{author}{\bibfnamefont{D.~M.} \bibnamefont{{Malioutov}}},
  \bibnamefont{and} \bibinfo{author}{\bibfnamefont{A.~S.}
  \bibnamefont{{Willsky}}}, \bibinfo{journal}{ArXiv e-prints}
  (\bibinfo{year}{2007}), \eprint{0710.0013}.

\bibitem[{\citenamefont{{B{\'e}ny}}(2013)}]{beny13}
\bibinfo{author}{\bibfnamefont{C.}~\bibnamefont{{B{\'e}ny}}},
  \bibinfo{journal}{ArXiv e-prints}  (\bibinfo{year}{2013}),
  \eprint{1301.3124}.

\bibitem[{\citenamefont{Saremi and Sejnowski}(2013)}]{Saremi19022013}
\bibinfo{author}{\bibfnamefont{S.}~\bibnamefont{Saremi}} \bibnamefont{and}
  \bibinfo{author}{\bibfnamefont{T.~J.} \bibnamefont{Sejnowski}},
  \bibinfo{journal}{Proceedings of the National Academy of Sciences}
  \textbf{\bibinfo{volume}{110}}, \bibinfo{pages}{3071} (\bibinfo{year}{2013}),
  \eprint{http://www.pnas.org/content/110/8/3071.full.pdf},
  \urlprefix\url{http://www.pnas.org/content/110/8/3071.abstract}.

\bibitem[{\citenamefont{{Miles Stoudenmire} and {Schwab}}(2016)}]{stoud16}
\bibinfo{author}{\bibfnamefont{E.}~\bibnamefont{{Miles Stoudenmire}}}
  \bibnamefont{and} \bibinfo{author}{\bibfnamefont{D.~J.}
  \bibnamefont{{Schwab}}}, \bibinfo{journal}{ArXiv e-prints}
  (\bibinfo{year}{2016}), \eprint{1605.05775}.

\bibitem[{\citenamefont{{Vidal}}(2008)}]{mera}
\bibinfo{author}{\bibfnamefont{G.}~\bibnamefont{{Vidal}}},
  \bibinfo{journal}{Physical Review Letters} \textbf{\bibinfo{volume}{101}},
  \bibinfo{eid}{110501} (\bibinfo{year}{2008}), \eprint{quant-ph/0610099}.

\bibitem[{\citenamefont{Ba and Caruana}(2014)}]{reallydeep}
\bibinfo{author}{\bibfnamefont{J.}~\bibnamefont{Ba}} \bibnamefont{and}
  \bibinfo{author}{\bibfnamefont{R.}~\bibnamefont{Caruana}}, in
  \emph{\bibinfo{booktitle}{Advances in neural information processing systems}}
  (\bibinfo{year}{2014}), pp. \bibinfo{pages}{2654--2662}.

\bibitem[{\citenamefont{Hastad}(1986)}]{hastad1986almost}
\bibinfo{author}{\bibfnamefont{J.}~\bibnamefont{Hastad}}, in
  \emph{\bibinfo{booktitle}{Proceedings of the eighteenth annual ACM symposium
  on Theory of computing}} (\bibinfo{organization}{ACM}, \bibinfo{year}{1986}),
  pp. \bibinfo{pages}{6--20}.

\bibitem[{\citenamefont{Telgarsky}(2015)}]{telgarsky2015representation}
\bibinfo{author}{\bibfnamefont{M.}~\bibnamefont{Telgarsky}},
  \bibinfo{journal}{arXiv preprint arXiv:1509.08101}  (\bibinfo{year}{2015}).

\bibitem[{\citenamefont{Montufar et~al.}(2014)\citenamefont{Montufar, Pascanu,
  Cho, and Bengio}}]{montu14}
\bibinfo{author}{\bibfnamefont{G.~F.} \bibnamefont{Montufar}},
  \bibinfo{author}{\bibfnamefont{R.}~\bibnamefont{Pascanu}},
  \bibinfo{author}{\bibfnamefont{K.}~\bibnamefont{Cho}}, \bibnamefont{and}
  \bibinfo{author}{\bibfnamefont{Y.}~\bibnamefont{Bengio}}, in
  \emph{\bibinfo{booktitle}{Advances in neural information processing systems}}
  (\bibinfo{year}{2014}), pp. \bibinfo{pages}{2924--2932}.

\bibitem[{\citenamefont{Eldan and Shamir}(2015)}]{eldan2015power}
\bibinfo{author}{\bibfnamefont{R.}~\bibnamefont{Eldan}} \bibnamefont{and}
  \bibinfo{author}{\bibfnamefont{O.}~\bibnamefont{Shamir}},
  \bibinfo{journal}{arXiv preprint arXiv:1512.03965}  (\bibinfo{year}{2015}).

\bibitem[{\citenamefont{{Poole} et~al.}(2016)\citenamefont{{Poole}, {Lahiri},
  {Raghu}, {Sohl-Dickstein}, and {Ganguli}}}]{poole16}
\bibinfo{author}{\bibfnamefont{B.}~\bibnamefont{{Poole}}},
  \bibinfo{author}{\bibfnamefont{S.}~\bibnamefont{{Lahiri}}},
  \bibinfo{author}{\bibfnamefont{M.}~\bibnamefont{{Raghu}}},
  \bibinfo{author}{\bibfnamefont{J.}~\bibnamefont{{Sohl-Dickstein}}},
  \bibnamefont{and}
  \bibinfo{author}{\bibfnamefont{S.}~\bibnamefont{{Ganguli}}},
  \bibinfo{journal}{ArXiv e-prints}  (\bibinfo{year}{2016}),
  \eprint{1606.05340}.

\bibitem[{\citenamefont{{Raghu} et~al.}(2016)\citenamefont{{Raghu}, {Poole},
  {Kleinberg}, {Ganguli}, and {Sohl-Dickstein}}}]{raghu16}
\bibinfo{author}{\bibfnamefont{M.}~\bibnamefont{{Raghu}}},
  \bibinfo{author}{\bibfnamefont{B.}~\bibnamefont{{Poole}}},
  \bibinfo{author}{\bibfnamefont{J.}~\bibnamefont{{Kleinberg}}},
  \bibinfo{author}{\bibfnamefont{S.}~\bibnamefont{{Ganguli}}},
  \bibnamefont{and}
  \bibinfo{author}{\bibfnamefont{J.}~\bibnamefont{{Sohl-Dickstein}}},
  \bibinfo{journal}{ArXiv e-prints}  (\bibinfo{year}{2016}),
  \eprint{1606.05336}.

\bibitem[{\citenamefont{Saxe et~al.}(2013)\citenamefont{Saxe, McClelland, and
  Ganguli}}]{saxe}
\bibinfo{author}{\bibfnamefont{A.~M.} \bibnamefont{Saxe}},
  \bibinfo{author}{\bibfnamefont{J.~L.} \bibnamefont{McClelland}},
  \bibnamefont{and} \bibinfo{author}{\bibfnamefont{S.}~\bibnamefont{Ganguli}},
  \bibinfo{journal}{arXiv preprint arXiv:1312.6120}  (\bibinfo{year}{2013}).

\bibitem[{\citenamefont{Bengio et~al.}(2007)\citenamefont{Bengio, LeCun
  et~al.}}]{bengio2007scaling}
\bibinfo{author}{\bibfnamefont{Y.}~\bibnamefont{Bengio}},
  \bibinfo{author}{\bibfnamefont{Y.}~\bibnamefont{LeCun}},
  \bibnamefont{et~al.}, \bibinfo{journal}{Large-scale kernel machines}
  \textbf{\bibinfo{volume}{34}}, \bibinfo{pages}{1} (\bibinfo{year}{2007}).

\bibitem[{\citenamefont{Strassen}(1969)}]{strassen1969gaussian}
\bibinfo{author}{\bibfnamefont{V.}~\bibnamefont{Strassen}},
  \bibinfo{journal}{Numerische Mathematik} \textbf{\bibinfo{volume}{13}},
  \bibinfo{pages}{354} (\bibinfo{year}{1969}).

\bibitem[{\citenamefont{Le~Gall}(2014)}]{le2014powers}
\bibinfo{author}{\bibfnamefont{F.}~\bibnamefont{Le~Gall}}, in
  \emph{\bibinfo{booktitle}{Proceedings of the 39th international symposium on
  symbolic and algebraic computation}} (\bibinfo{organization}{ACM},
  \bibinfo{year}{2014}), pp. \bibinfo{pages}{296--303}.

\bibitem[{\citenamefont{Carleo and Troyer}(2016)}]{carleo}
\bibinfo{author}{\bibfnamefont{G.}~\bibnamefont{Carleo}} \bibnamefont{and}
  \bibinfo{author}{\bibfnamefont{M.}~\bibnamefont{Troyer}},
  \bibinfo{journal}{arXiv preprint arXiv:1606.02318}  (\bibinfo{year}{2016}).

\bibitem[{\citenamefont{Vollmer}(2013)}]{vollmer2013introduction}
\bibinfo{author}{\bibfnamefont{H.}~\bibnamefont{Vollmer}},
  \emph{\bibinfo{title}{Introduction to circuit complexity: a uniform
  approach}} (\bibinfo{publisher}{Springer Science \& Business Media},
  \bibinfo{year}{2013}).

\end{thebibliography}

\end{document}